# THE THOUSAND STAR MAGNITUDES IN THE CATALOGUES OF PTOLEMY, AL SUFI, AND TYCHO ARE ALL CORRECTED FOR ATMOSPHERIC EXTINCTION

BRADLEY E. SCHAEFER, Louisiana State University

## 1. PRE-TELESCOPIC CATALOGUES

Three pre-telescopic star catalogues contain about a thousand star magnitudes each (with magnitudes 1, 2, 3, 4, 5, and 6), with these reported brightnesses as the original basis for what has become the modern magnitude scale. These catalogues are those of Ptolemy (*c.* 137, from Alexandria at a latitude of 31.2°), Al Sufi (*c.* 960, from Isfahan at a latitude of 32.6°), and Tycho Brahe (*c.* 1590, from the island of Hven at a latitude of 55.9°). Previously, extensive work has been made on the *positions* of the catalogued stars, but only scant attention has been paid to the *magnitudes* as reported[1]. These magnitudes will be affected by a variety of processes, including the dimming of the light by our Earth's atmosphere (atmospheric extinction), the quantization of the brightnesses into magnitude bins, and copying or influence from prior catalogues. This paper provides a detailed examination of these effects. Indeed, I find all three catalogues to report magnitudes that have near-zero extinction effects, so the old observers in some way extinction corrected their observations.

The ancient star catalogue of Ptolemy appears in Books 7 and 8 of the *Almagest*, with positions and magnitudes for 1028 stars. These magnitudes are in the now-traditional system of 1, 2, 3, 4, 5, and 6, however with notations for stars that are somewhat brighter or fainter than the integral magnitude. Thus, the notations go, from the nominal brightest to faintest, 1, <1, >2, 2, <2, >3, 3, <3, >4, 4, <4, >5, 5, 6, and faint. A handful of stars are duplicates or marked as nebulous instead of being given a magnitude. I adopt the magnitudes and star identifications as given in the translation of G. J. Toomre[2]. Ptolemy does tell us how to measure star positions using armillary spheres, but he does not give one word on how the magnitude scale was set nor how to measure magnitudes. There is a substantial long-running debate as to whether the *Almagest* star catalogue was primarily observed by Ptolemy or rather Hipparchus (*c.* 128 B.C. from Rhodes with a latitude of 36.4°)[3]. This debate has a wide variety of arguments running many levels deep, with neither side able to produce decisive evidence to convince the other side. The primary material for this debate is the star positions recorded in the catalogue, with scant use having been made of any of the concurrent magnitude information[4].

In the western world, the next star catalogue came from the Persian astronomer 'Abd al-Rahman al-Sufi (903-986) observing from Isfahan (latitude 32.7°). In his *Book of Fixed Stars*, published in 964, Al Sufi's star list gives the same stars and star positions (updated for precession) as in the *Almagest*. However, Al Sufi observed his own magnitudes, and these are substantially different from those in the *Almagest*. I adopt the magnitudes and star identifications as given for Al Sufi by E. B. Knobel[5]. Al Sufi gives magnitudes in the same basic system as the *Almagest*, with the notation for the brightness bins, nominally from brightest to faintest, being 1, 1-2, 2-1, 2, 2-3, 3-2, 3, 3-4, 4-3, 4, 4-5, 5-4, 5, 5-6, 6-5, 6, and 6-7. These are one-third magnitude bins, with several extra when compared to the *Almagest* (5-6, 6-5, and 6-7). To say 'one-third magnitude bins' is approximately correct, but it is really a schematic description for categories that are variable in size with ill-defined edges and imperfect measurements. Nothing survives which tells us the details of how Al Sufi measured his magnitudes.

The next star catalogue is that of Ulugh Begh (1394-1449), the grandson of Tamerlane, who ruled a large region of central Asia from Samarkand. He noted errors in the positions of the stars from the *Almagest*, collected a group of scholars, and a star catalogue was made from observations in Samarkand (latitude $\lambda$=39.7°) around the year 1437[6]. His star catalogue contains the same stars as are in the *Almagest*. The positions of the stars were newly measured with large sextants, armillary spheres, and meridian circles. But the magnitudes were copied from Al Sufi, with all of the magnitudes rounded to the nearest integer. (For example, Al Sufi's stars labeled 3-2, 3, and 3-4 were all labeled as 3 by Ulugh Begh.) As the magnitudes are simply copied from Al Sufi, this star catalogue will not be considered further here.

The last pre-telescopic star catalogue (in the western world) was observed by Tycho Brahe (1546-1601) from his island of Hven at latitude 55.9°. The star positions and their magnitudes were measured from 1589 to 1591, from which a catalogue of 777 stars appeared at the end of 1592[7] as published in his *Progymnasmata*[8]. Just 28% of these stars have notations (either a colon or a period following the integer) that indicate that the observed magnitude is somewhat brighter or dimmer, respectively, than the integral magnitude. So for example, stars labeled "3:", "3", and "3." have average modern V-band magnitudes of 2.88, 3.25, and 3.62 mag respectively. From Tycho's *Progymnasmata*, we are given substantial details on how he measured the positions of the stars with large scale meridian circles and sextants. However, I know of no place in which Tycho tells us about how he measured the magnitudes[9]. From 1595 to 1597, Tycho restarted the observing so as to bring his number up to a thousand stars, although these observations were hasty and at least the positions have substantial problems[10]. The resulting 1004 star catalogue has appeared several times[11], with the magnitudes now only given to the nearest integer. As with the other catalogues, Tycho's catalogue has been corrected and the star identifications discussed and improved, but all of these small variations make no significant difference to the work in this paper, because I am operating off a large number of stars so that any small number of remaining mis-identifications are negligible. Nevertheless, the provenance and numbers differ substantially in the two versions of the catalogue, so I will treat the 777-star and 1004-star versions with parallel independent analyses.

## 2. MAGNITUDES AND EXTINCTION

The star brightnesses reported in the old catalogues are on a scale of magnitudes, where the brightest stars are first magnitude, the next group of stars are of the second magnitude, and so on down to sixth magnitude. The earliest known appearance of this system is in the *Almagest*. This magnitude system was followed by all subsequent western works and formed the original basis for the modern magnitude system. Appendix 1 gives much further detail on the magnitude system.

To measure a star's magnitude, the only way to do this is to somehow compare its brightness with some other star(s) of stated magnitude(s) or some stated standard threshold. I can think of many plausible variations by which the old magnitude system could have been defined. For an example of a hybrid system that could have worked, the first magnitude stars were taken to be the brightest dozen or so, the sixth magnitude stars are those that are just barely visible under clear dark skies, while the intermediate stars are taken to be those that closely match some set of standard stars (like Polaris being the definition of a second magnitude star). We have no guidance from old sources as to how the scale was originally defined, either in theory or practice. Nevertheless, all measurement methods must ultimately compare the star's observed brightness against some standard or standards.

One problem is to compare the old magnitude systems with modern magnitudes. This is a problem because the old magnitudes are not exactly equal to, nor even linearly dependent on, the modern magnitude scale. The brighter stars are generally pretty close, but the faintest stars are reported to be fainter than the modern measures. For example, stars labeled by Ptolemy as "6" have an average modern magnitude of V=5.20. And the relationship is not even monotonic, with the stars labeled "<4" actually being fainter on average than those labeled ">5", "5", and "faint". We can only accept the old magnitudes as being binned with some average that must be empirically determined. So, when Ptolemy says that a star is sixth magnitude, then we should interpret this as a report that the star was approximately m=5.20. All the stars labeled as sixth magnitude have a substantial scatter around this average (with an RMS scatter of 0.37 mag in this case), so we would really take the sixth magnitude star to have a reported brightness of m=5.20±0.37. In Appendix 1, I have tabulated the average modern magnitudes for all star brightness labels for all three catalogues. With this, we can get the modern equivalents of each reported observation, with these serving to translate the old reported brightness into the best estimate for what that report means.

These translated old magnitudes can be designated as 'm', while the same star has a modern V-band magnitude designated as 'V'. The deviation between the reported magnitude and the modern magnitude is m-V. In an ideal world with perfect observations, m-V=0. With the inevitable scatter due to observational errors (typically one third of a magnitude) and the quantization of the magnitudes into bins, the values of m-V will differ from zero, typically with an RMS scatter of around half of a magnitude.

The average m-V will equal zero, by construction, for the stars used to find the bin average. For the many cataloged stars, the measured quantity m-V will provide a measure of deviations from the average, for example, resulting from atmospheric extinction.

Another complexity is that the atmospheric extinction will dim the light of the stars. That is, light passing through the Earth's atmosphere will be scattered (by Rayleigh scattering off gas molecules and by Mie scattering off aerosols) and absorbed (by ozone in the stratosphere) so that the star will appear dimmer than if there was no atmosphere. When the star appears at the zenith, one pathlength of atmosphere (called one airmass) will dim the star by roughly a third of a magnitude for good clear skies. For stars that appear away from the zenith, the light will pass through a pathlength of air that is X airmasses thick. When stars and their comparison stars are both near the zenith, the dimming equally affects both stars, so the comparison remains the same. But when a star is 60° from the zenith, the starlight has to pass through twice as much air as at the zenith (with X=2.00), so a third of the incoming light is lost in the first airmass and then another third of the remainder is lost for the second airmass, so the star looks just under half as bright as compared to if it were viewed at zenith. When the star gets low towards the horizon, the starlight passes through many airmasses and appears greatly dimmed. This dimming depends on the zenith distance of the star and on the haziness of the atmosphere. The haziness of the atmosphere is quantified by a parameter called the extinction coefficient, $k$, with units of magnitudes lost per airmass. Typical values for near-sea-level sites around Europe and the Mediterranean are 0.23 mag/airmass for the best nights to ~0.4 for average nights, with a selection of the better nights giving perhaps 0.25 mag/airmass.

The basic physics of atmospheric extinction was first presented by Pierre Bouguer[12] in 1729, and this analysis is essentially the modern understanding. Previously, in 1723, Jean-Jacques de Mairan recognized the existence of extinction, but his understanding was vague and his explanation was faulty[13]. For any time before 1723, I know of no analysis, nor any prior mention of the phenomenon of extinction[14]. In Appendix 2, I give the modern model for extinction, along with an explicit equation for the dimming of the light (m-V) as a function of the star's zenith distance and $k$.

Stars that are anywhere near overhead suffer only negligible dimming, m-V≈0, but the fading increases rapidly as the horizon is approached. Thus, for an adopted extinction of 0.25 mag/airmass, the dimming, m-V, is 0.00 mag at the zenith, 0.025 mag for 25° from the zenith (and is completely unobservable), 0.25 mag for 60° from the zenith (30° above the horizon), 2.3 mag for 85° from the zenith (5° above the horizon), and around 10 mag for stars at the horizon (making all stars invisible).

The stars that pass overhead (say, those with declinations from +6° to +56° as viewed from Alexandria) will suffer negligible extinction around culmination. The most likely and natural time to make observations of stars is when they are within a few hours of culmination. This is not a requirement, but this is so natural that this must be taken as the default, while the making of magnitude estimates far from culmination has no precedent. The dimming changes little for long times on either side of culmination, so there is a very wide window for the effects to be negligible. For example, a star that passes overhead near the zenith has a time interval of 7.0 hours during which it can be observed with less than 0.1 mag of dimming (for k=0.25 mag/airmass). For a star that culminates 10° above the horizon, there is a 100 minute interval centered on culmination over which the dimming is within 0.1 mag of the dimming at culmination. Thus the idealization that the magnitude measures where made with the star at culmination is actually a very good approximation for a very wide window of observing times.

For catalogued stars north of the celestial equator, their measured magnitudes will be when both they and their comparison stars are in the large region of negligible extra dimming. For these stars, we expect m-V≈0. But for stars with far southerly declinations, those that never rise high above the southern horizon, they are always seen through high airmass and hence will always appear greatly dimmed. Stars culminate with a zenith distance of $|\lambda-\delta|$, where $\lambda$ is the observer's latitude and $\delta$ is the star's declination. From Alexandria, a star with declination -29° will be dimmed by m-V=0.25 mag when compared to comparison stars near the zenith. This size of an effect is too small to be confidently noticed for any one star, but we have of order a hundred stars of comparable declination and then the random observational errors (due to ordinary measurement errors plus the quantization into magnitude bins) will be beaten down to where the effect is easily noticeable. For a star with $\delta$=-48.8° (culminating 10° above the horizon from Alexandria), the dimming is large, with m-V=1.16 mag, and should be noticeable even for individual stars. The point of this analysis is that the southerly stars were observed significantly dimmed compared to their modern magnitudes, and this dimming has m-V going as a sharply rising function as the declination approaches the southern horizon. This is the key analysis point of this paper.

We can measure the m-V dimming for each catalogued star (see Appendix 1). Then we can see how the dimming changes with the star's declination. In particular, we can fit the observed m-V for the thousand stars to the model m-V (see equation A4). This model is only a function of the observer's latitude ($\lambda$) and the extinction coefficient (*k*). With the standard chi-square fitting procedure, we can derive the best fit $\lambda$ and *k* values, as well as their one-sigma error bars. Appendix 2 gives the model equation, examples of how m-V varies with declination and extinction, as well as a simulated data set to illustrate the process. Importantly, for the real catalogues with a thousand stars, the observer's latitude can be determined to much better than one degree of latitude. This accuracy is confirmed to be easy to attain with intentionally-casual modern observations using zero equipment and zero modern theory (see Appendix 4).

My original idea was to determine the latitude of the observer of the *Almagest* magnitudes, with this presumably sharply distinguishing between Ptolemy in Alexandria

(λ=31.2°) and Hipparchus in Rhodes (λ=36.4°).  As we shall see, this goal was not achieved, for a completely unexpected reason.

## 3. THE *ALMAGEST* STAR CATALOGUE

The m-V values (the atmospheric dimming plus observational scatter) can be calculated for all the stars in the *Almagest* as described in Appendix 1.  Over most declinations, the extinction is small and negligible.  But for declinations around -30° (with nearly X=2 and a dimming of around a quarter or a third of a magnitude), the stars should be noticeably dimmer in their reported magnitudes than expected, with this being clearly measurable for the many catalogued stars  For the most southerly stars, never seen from Alexandria at higher than 10° above the horizon, the stars will always appear more than one magnitude fainter than expected, with this apparent even on a star-by-star basis.  To quantify this, I have plotted m-V versus declination in a standard format, where we should see a scatter of roughly half a magnitude (due to the usual photometric scatter plus the quantization of the magnitudes) superposed on some sort of an extinction curve.  Figure A2 presents a simulated data set for the *Almagest* so that we can see what is expected.

The *Almagest* m-V versus declination curve is presented in Figure 1.  We see a scatter of nearly half a magnitude, as expected, plus a somewhat larger scatter towards the southernmost stars.  The immediately striking point from this figure is that the stars do *not* flare up at the leftmost edge.  That is, the southernmost stars are *not* reported to be much dimmer than expected.  This is startling and without precedent.

I have made a formal chi-square fit to the data in Figure 1 and found $k=0.010\pm0.017$ mag/airmass.  This is greatly different from the expected $k\approx0.25$ mag/airmass, and is zero to within the small error bars.  (With zero extinction, it does not matter what the latitude is taken to be.)  It is impossible that the real extinction could have been that low, because the molecules in our air provide 0.14 mag/airmass of extinction at sea level while omnipresent aerosols are never less than something like 0.1 mag/airmass at sea level sites in pre-industrial times.  This essentially zero extinction is the central dilemma of this paper.

This apparently zero extinction is not just the result of a few southernmost stars.  The effect is smaller for more northerly stars, but there are many more of them in the catalogue.  We can see the same result if we restrict ourselves to only looking at stars north of declination -48.8° (so that they all culminate more than 10° above the southern horizon of Alexandria).  For this case, the fitted *k* is $-0.010\pm0.021$, which is not significantly different.  And the effect can also be seen and proven with only middle altitude stars.  So, for the 110 stars that culminate with altitudes of 20° to 30° from Alexandria (between declinations -38.8° and -28.8°), the average m-V is -0.043 mag.  The RMS scatter of m-V for these 110 stars is 0.73 mag, so the uncertainty in the mean is ±0.070 mag.  These middle-altitude stars with average m-V of -0.043±0.070 mag are consistent with perfect extinction correction (which predicts the average m-V to be zero). For these stars (with an average altitude of 25° and X=2.4) and extinction of 0.25 mag/airmass, the dimming should be by 0.34 mag. So the reported magnitudes of the 110 middle-altitude stars (with average m-V of -0.043±0.070 mag) are inconsistent with the no-correction case (with average m-V of +0.34 mag) at the 5.5-sigma level. This can even be pushed to more northerly stars, for example the 128 stars from declination -25° to -15° have average m-V of -0.057±0.059 (consistent with extinction correction), while being inconsistent with the expected dimming (0.15 mag) at the 3.5-sigma level. So the basic result (the southerly stars are not reported to be dimmed) is very robust, as it holds for all stars from the most southerly to the ones that culminate halfway towards the zenith.

One immediate consequence of this result is that my original goal (deriving the latitude of the observer of the magnitudes) cannot be reached. With near-zero effective extinction, the effects of changing the latitude are near-zero, so no distinction can be made between Ptolemy and Hipparchus as based on magnitudes alone. Instead, I have found a completely unexpected result that is highly intriguing in its own right.

Figure 1 does not display the expected effects of extinction, and the extinction was certainly present when the observer was measuring the star magnitudes. So somehow, the reported magnitudes have been corrected for extinction. There are a variety of ways in which this extinction correction could have arisen. Perhaps the observer (consciously or unconsciously) made some sort of a crude and empirical correction from the observed magnitude to the reported magnitude. (Such modifications need no physical understanding or sophistication.) Or perhaps the observer used some procedure that happened to produce effectively extinction corrected magnitudes, where there would be no need to even know about the existence of extinction. We have no guidance from historical sources, and I can think of many possibilities. So the question of how the reported magnitudes turn out to be extinction corrected is the central mystery of this paper.

An important clue relates to how the extinction correction varies along the southern horizon. For the *Almagest*, the *average* extinction is basically zero, with this resulting from southern stars strung out with over all right ascensions (RA). However, we know that the *selection* of stars along the southern horizon was distinctly different for the RA range 0°-270° as compared to the 270°-360° quadrant[15], so it is reasonable to look for variations of the effective extinction coefficient with ranges of RA. Indeed, I find large and highly significant variations of the fitted $k$ values for ranges of RA. For RA from 270° to 45° (passing through 0°), I find the best fit $k$ value of -0.30±0.05 mag/airmass. A negative $k$ value is not physical, even though it can be calculated within the extinction model. This value of $k$ is like the reported magnitudes were *double corrected* for extinction. For RA from 135° to 225°, I find the best fit $k$ value of 0.165±0.027 mag/airmass. This is like the reported magnitudes were *half corrected* for extinction. For the two intermediate ranges of RA (45°-135° and 225°-270°), the extinction appears to be in a transition that is not resolved. Thus, from Capricorn to Fomalhaut to Eridanus, the effective extinction applied is 2X too large, while from the middle of Hydra to Centaurus to Scorpius, the effective extinction applied is 2X too small.

The plot of m-V versus declination for stars within these two RA ranges show a downturn to the left (for 270° to 45°) and an upturn to the left (for 135° to 225°). In both these cases, the scatter for the southernmost stars is comparable to the scatter for the stars that pass overhead. Figure 1 is a superposition of these two plots (plus the transition regions), and the net result of the upturn and downturn is that the average is flat. That is, the local corrections for extinction are far from perfect, and it is only happenstance that the corrections average out very close to zero. On the left side of Figure 1, the superposition of the upturn and downturn results in the apparent increase in the scatter to the left.

The *Almagest* stars far to the south must have been actually seen to the observer to be substantially dimmed with respect to comparison stars higher up in the sky, but the southern stars are *not* reported to be dimmed at all (on average). The effective extinction varies from -0.30 to +0.165 mag/airmass along the southern horizon, even though the overall average is essentially zero (0.010±0.017 mag/airmass). This is startling.

## 4. THE AL SUFI STAR CATALOGUE

Similarly, the m-V values can be calculated for every star in the catalogue of Al Sufi, and then plotted against the declination of the star for the epoch 964 AD (see Figure 2). Again, we do not see the expected sharp upturn towards the left. This immediately shows that the reported star magnitudes must have been somehow measured or modified such that the extinction effects are largely taken out. That is, the magnitudes in the Al Sufi star catalogue are extinction corrected in some unknown way.

A formal chi-square fit to all the stars (for the adopted latitude of 32.6°) gives k=+0.057±0.007 mag/airmass. This is significantly different from zero, but is far smaller than any plausible value for the real extinction of the atmosphere. It is like some empirical correction or procedural method accounted for only ~75% of the extinction. Or more likely, ordinary errors in this empirical correction or procedural method made for imperfect extinction corrections, and in this case the value used was only 25% off. In Appendix 4, I report how my modern measurement of the extinction correction was a similar degree off, so this is plausible. Just as with the *Almagest*, we have the case where Al Sufi reports star magnitudes for his southern stars to be substantially brighter than they must have appeared to the observer, which is to say that the magnitudes are somehow extinction corrected.

Again, a critical clue as to the nature of the correction comes from the variations of the fitted $k$ value for different regions of stars along the southern horizon. For the RA range 270° to 45°, the extinction is -0.012±0.041 mag/airmass. For the RA range of 135° to 225°, $k$ equals 0.052±0.008 mag/airmass. Both of these values are greatly different from those in the *Almagest* for identical RA ranges. Nevertheless, the fitted $k$ values change greatly along the southern horizon, with $k$ equal to -0.090±0.048 mag/airmass in the first quadrant (0°-90°) with relatively few stars far to the south, while $k$ equals +0.095±0.011 mag/airmass in the third quadrant (180°-270°). This pattern is greatly different from the *Almagest*.

Al Sufi's star catalogue copied the positions of the stars from the *Almagest* along with a correction for the effects of precession, so it is reasonable that he might have also copied all or part of the magnitudes. However, we quickly see that Al Sufi's magnitudes are often different from those of Ptolemy, so they are not simply copied. And indeed, Al Sufi has some magnitude bins that are not included by Ptolemy. In Appendix 3, I make a thorough analysis of possible copying. Only 55% of Al Sufi's magnitudes are identical to those reported in the *Almagest*, and this sets an upper limit on the copying fraction. For stars within a small range of V magnitudes, the number of identical reported magnitudes is substantially higher than if the Al Sufi reports were uncorrelated with those of Ptolemy. And the number of stars with largely discrepant magnitudes in the *Almagest* (with m-V more than one magnitude from zero) have a large fraction in Al Sufi's book with similar large discrepancies. In all, I conclude that Al Sufi copied roughly a third of his magnitudes and was greatly influenced for another third of his magnitudes.

So a substantial fraction of Al Sufi's magnitudes are *not* independent of Ptolemy, and as such the extinction correction in Figure 2 is partly a simple legacy from the *Almagest*. But this is not the whole story as we can ask whether the magnitudes observed by Al Sufi are extinction corrected. For this analysis, I have selected out only those stars whose magnitude differs from Ptolemy greatly, such that the two magnitudes do not share the same integer, for which Al Sufi neither copied nor was influenced by Ptolemy. For example, if Ptolemy reports a star as >4, then Al Sufi *might* have copied the magnitude if he also reported the magnitude as 4-3, while Al Sufi might have been influenced in his magnitude if he reported 4 or 4-5, but if Al Sufi reports a magnitude of 3, 3-4, or 5-4 then it is certainly independent. A total of 164 such stars were definitely observed by Al Sufi. For these, the fitted extinction is 0.072±0.017 mag/airmass. This value is consistent with the overall fitted extinction, is significantly different from zero, and is significantly smaller than any physically possible value for extinction. There are too few southerly stars to well-resolve the behaviour along RA ranges, but the same pattern as for the totality of stars (in particular the relatively high extinction in the third quadrant) is apparently repeated. So the star magnitudes certainly observed by Al Sufi are definitely extinction corrected (although not perfectly).

I have also examined the subset of 454 stars for which Al Sufi and Ptolemy quote different magnitudes. This subset certainly excludes all copied magnitudes, although there may be Al Sufi magnitudes that have been influenced by Ptolemy. The fitted value is k=0.075±0.017 mag/airmass, which again proves that the magnitudes as measured by Al Sufi are extinction corrected. The pattern with RA is greatly different from that of the *Almagest* and is the same as for the greatly-different Al Sufi stars.

Al Sufi has 551 stars with magnitudes identical to those given in the *Almagest*, and this subset of stars includes copied magnitudes, influenced magnitudes, and independently measured magnitudes. If these were all copied from Ptolemy, then I would expect to see the same fitted *k* values and the same pattern with RA as in the *Almagest*. If these were all independent, then I would expect to see the same values and patterns as for the Al Sufi stars with greatly different magnitudes. Instead, the fitted *k* value for all the stars in this subset (0.049±0.009 mag/airmass) is roughly halfway between the two extremes. This suggests that only half of this subset is copied or influenced. And the pattern in the *Almagest* (with k=+0.165±0.027 for 135°<RA<225°) is not seen as Al Sufi's stars give k=+0.042±0.009 mag/airmass. This result suggests that Al Sufi's copy fraction is substantially lower than indicated from the analysis in Appendix 3. A way to reconcile the above results with Appendix 3 is to think that Al Sufi only copied those magnitudes from the *Almagest* when they were similar to Al Sufi's own estimate.

In all, we find that the magnitudes reported by Al Sufi had an imperfect extinction correction (roughly 3/4 of what the correction should have been), and something like one-quarter to one-half of his magnitudes were copied from or influenced by Ptolemy.

## 5. TYCHO'S STAR CATALOGUE

Tycho has two versions of his star catalogue, the early 777-star version and the later 1004-star version. I have calculated the m-V values for each of the entries in these catalogues (see Appendix 1), and I have plotted these values versus the star declinations in Figures 3 and 4. Again, we see that the plots have no upturn to the left. The fitted *k* values are 0.044±0.011 mag/airmass for the 777-star catalogue and 0.013±0.012 mag/airmass for the 1004-star catalogue. I have also picked out the added stars (nominally with 1004-777=227 stars), and fitted these to find *k* equal to -0.069±0.029 mag/airmass. A key new result of this paper is that both the plot and the formal fit show that Tycho's magnitudes are extinction corrected with pretty good accuracy.

In Appendix 3, I evaluate the fraction of magnitudes in Tycho's catalogue that were copied or influenced from the *Almagest*. This fraction is found to be small, perhaps 10% or perhaps even zero.

I have also sought variations in the *k* value for southern stars over various ranges of RA. For the 1004-star catalogue, the four quadrants have *k* values of -0.025±0.038,

+0.055±0.025, -0.001±0.022, and +0.028±0.017 mag/airmass respectively, which shows that the extinction does not vary significantly change with RA.  For the ranges from 135°-225° I find *k* equal -0.065±0.035 mag/airmass, and from 270°-45° I find *k* equal +0.001±0.017 mag/airmass.  This is completely different from the *Almagest* pattern.  The same conclusions hold strongly for the 777-star catalogue.

However, there are two localized anomaly that appears only in Tycho's catalogues.  The first anomaly appears only in the stars added to the 777-star catalogue to make the 1004-star catalogue.  The average m-V value from declination -20° to +30° is 0.31±0.04 mag more negative than the stars from the fiducial declination region of +32° to +80°.  That is, for some unknown reason, Tycho's added stars in the 'tropical region' (including the zodiacal stars) are systematically reported to be *brighter* than expected.  This has nothing to do with extinction, but likely has everything to do with how Tycho set up the calibrations for his later additions.  This anomaly appears to be consistent throughout the range of RA, and the anomaly does not appear in the 777-star catalogue.

The second localized anomaly appears only in the first 777-star catalogue, and not in the added stars.  The southern stars for RA range 270° to 310° (Sagittarius and Capricorn) have a sharp upturn in m-V to low declinations.  With little scatter, the m-V value increases roughly linearly from close to 0.00 mag at $\delta=-17°$ to +1.5 mag at $\delta=-28°$.  The shape of this rise is distinctly different from that expected for any simple extinction effect.

Including the four anomalies from the *Almagest* and Al Sufi (for 135°<RA<225° and 270°<RA<45° by Ptolemy and the first and third quadrant by Al Sufi), we now have six localized anomalies in the effective *k* values.  I interpret these anomalies to be caused by large systematic errors as a function of altitude in the extinction correction method, with these errors being applicable only over some restricted region of the sky.  For example, perhaps the extinction correction in Tycho's catalogue was made by an informal instinctive correction estimated by the observer (see next section), and the Sagittarius/Capricorn region happened to have been observed by one of Tycho's assistants with a different perceived correction.  Or perhaps Tycho was making monthly measures of some extinction correction table (like he repeatedly measured for refraction tables), and the Sagittarius/Capricorn estimates happened to be the only southern stars for which ordinary fluctuations in the apparent extinction happened to produce a monthly table with very low corrections.  The fact that these anomalies are not mirrored from the *Almagest* to later catalogues proves that the extinction corrections in the later catalogues are not simple products of Ptolemy's extinction corrections.  The fact that these anomalies are localized over the sky and change from early to late times proves that the extinction correction method is not monolithic, but instead is from some method/procedure that is repeatedly applied with changing output.

# 6. HOW TO CORRECT THE MAGNITUDES FOR EXTINCTION?

The magnitudes reported for the southern stars are *not* dimmed as expected for extinction, so they are somehow extinction corrected. The central question of this paper is to explain this basic result. Unfortunately, we have no old texts that tell us about extinction or its corrections, so we are left with using indirect evidence. There are many possible ways to correct for the extinction. The two basic paths involve the observer taking the observed magnitude and modifying it on some basis, or the observer using some procedure that automatically makes the extinction correction. However, it is easy to spin further ideas to explain the extinction correction, so I will discuss additional hypotheses along with their refutations.

### 6.1. *Extinction correction by modifying the observed magnitude*

A modern astronomer would produce a catalogue of star magnitudes by observing their brightness (as dimmed by extinction) and then correct the observed magnitudes by means of some calculation based on a model. For this, the observed brightnesses would be compared back to the brightnesses of standard stars of known magnitude, and the extinction *k* would be explicitly measured from standard star measures. However, this modern approach is certainly an anachronism[16], completely inappropriate for the times and knowledge of any of the pre-modern observers.

Nevertheless, it is fully possible for a pre-telescopic observer to modify their observed magnitudes (intentionally or unintentionally) so as to get a perhaps crude and empirical correction. In its simplest form, the observer has long experience with watching stars rise and set and has some sort of an internal sense as to the amount of extinction as a function of the star's altitude, and then the observer, consciously or unconsciously, changes the observed magnitude to the reported magnitude. Such an informal and approximate method would need no modern sophistication or equipment, and it does not even need any conscious realization of the extinction phenomenon[17]. For a perceptive observer, the estimation task might go something like seeing the brightness of the southern star, recalling roughly the dimming at that altitude from watching more northerly stars rise/set, then bumping up the reported magnitude into the next bin if the star appears on the edge. This extinction correction method is unsatisfactory because it is hard to test, but test-worthiness is not a criterion for evaluating validity. A potential problem is whether such an estimation can correct for extinction, on average, such that the result has k≈0? Nevertheless, this idea is plausible and possible.

There is a middle ground between the modern sophistication and the casual modification. It is fully possible and plausible for the observers to systematically organize a quantitative or qualitative correction for the observed magnitudes. Appendix 4 gives full details of one such method that uses zero equipment and zero modern knowledge. Indeed, my recent trials of this method show that even casual attempts at correction can quickly and easily produce corrections that are comparable in accuracy to those observed for the catalogues. This modern trial does not prove that this is the method of the observers, yet it does prove that some systematic correction is easy and possible.

A trouble for this middle path (involving the creation of something like a correction table, graph, or equivalent) for the catalogue of Tycho is that we see no such calculations reported in his *Opera Omnia*. For example, we only see recorded the simple statement of magnitudes, and never any rise/set observations, tables, or graphs. I do not view this as a serious trouble because many of Tycho's positional calculations and corrections are also lost, so the lack of records is easy explained as also being lost.

An important question is whether such corrections are anachronisms? The informal correction idea is the sort of adjustment that humans are making all the time, so this possibility is certainly not an anachronism. But systematic and empirical correction needs the will to improve the reported magnitudes, the culture of making corrections, and possibly the culture of being quantitative. However, these requirements are exactly what we know already was the case for the three old observers. Tycho made quantitative and detailed measurements for the purpose of making refraction corrections in the observed altitude of stars in his catalogue. Al Sufi explicitly made quantitative and detailed corrections for the precession of star longitudes from those given by the *Almagest*. Ptolemy made additions of epicycles in an explicit quantitative correction so the model would match the positional observations. All three observers were frequently in the habit of making quantitative corrections, with this being a driving force for them and in their culture. So the making of extinction corrections is certainly not an anachronism.

### 6.2. *Extinction correction by use of some procedure*

Perhaps the observers used some procedure that automatically corrected for extinction, with this being either intentional or unintentional. We are told nothing about their methods for measuring magnitudes, so we can only speculate about what procedures could result in extinction corrections. I can think of only one general procedure that would produce any sort of automatic extinction correction, and that is to observe all stars with a closely similar altitude as their comparison stars. For example, Canopus can be compared to Arcturus and Spica when they have risen to 6° above the horizon, and the magnitudes of these comparison stars can be calibrated when they are high overhead. I can think of two specific procedures that will automatically make for extinction corrected magnitudes.

The first specific procedure is simply for the observer to await a time when his comparison star has nearly the same altitude as his target star, then make his magnitude estimate by a direct comparison. In this case, the extinction will be similar for the target and the standard, so the differential extinction will be small, and the target star will be estimated without any effect due to extinction.

Let me give an example of this procedure, highlighting the effort and the precision required to get an accuracy of roughly half a magnitude. Let us consider what it takes to make magnitude estimates for Canopus and the nearby stars in Puppis. For Canopus ($\alpha$ Car at V=-0.70 and $\delta$=-52.5°), the only first magnitude stars that rise or set while Canopus is visible are Arcturus and Spica. Both of these comparison stars are on the equator or further north, so they pass enough overhead that they can be calibrated when overhead with negligible extinction corrections. (Fortunately, I cannot think of any occasion when a two-step calibration would be required.) For Alexandria, Arcturus is rising when Canopus has an altitude of 6.2° and m-V=1.91 (for k=0.25 mag/airmass). To have less than a one-third of a magnitude error (i.e., so 1.58<m-V<2.24), the altitude of Arcturus must be between 5.2° and 7.5°. This corresponds to a time interval of 12 minutes. Observations made outside this rather sensitive time interval will produce a larger than a third of a magnitude error which (when added in quadrature with the other errors) will result in an increased scatter in m-V, with this not being observed for the southerly stars. The implication is that the observer would have to spend a lot of time waiting for their comparison star to get to exactly the right altitude, they must have some means to accurately estimate altitudes, and they must have used this means frequently throughout the waiting time. And such must be done for each star individually. If this procedure was adopted, the observer must have added a deep layer of complexity that required very large time and effort. The nearby stars in Puppis ($\sigma$ Pup at V=3.25 and $\delta$=-40.6°, $\nu$ Pup at V=3.17 and $\delta$=-42.8°, and $\tau$ Pup at V=2.93 and $\delta$=-49.4°) will only require one comparison star (although more would be better) at V≈3. Such a comparison star might be Vindemaitrix ($\epsilon$ Vir at V=2.83 and $\delta$=21.3°). Observations of the stars in Puppis all have substantially different declinations, so the observer would have to time their estimates carefully, with each star at a different time even though they are all using the same comparison star. With this procedure involving choosing a comparison star and waiting for just the right time to make an estimate, it is clear that relatively few stars can be estimated in any one night.

    Canopus and the nearby stars in Puppis can be observed with a minimum of two comparison stars (say, Arcturus and Vindemaitrix). The next constellation over will require a completely different set of 2-5 or more comparison stars. And each southern constellation will require a set of comparison stars that is along a unique arc in the northern skies. Thus to map out the southern star magnitudes, the observer must have a previously calibrated set of ~50 northern stars spread out roughly uniformly around the sky. The magnitude measures would have to proceed in at least two steps, where the northern comparison star network are given their estimated magnitudes, and then to use these magnitudes at carefully chosen small time intervals to estimate the southern star magnitudes.

    This first specific procedure is conceptually simple and effective, so this idea will be alluring to armchair astronomers. But this procedure is also completely un-

natural, tedious, awkward, and incredibly slow, so that no observer would ever think to do it unless they are explicitly trying to correct for extinction. Thus, to accept this specific procedure, we would have to think that the observer knew about extinction and was performing a huge amount of extra work so as to correct for it with this simple method.

The second specific procedure is another variant on the basic idea of estimating the target star magnitudes at the same altitude as the comparison stars. In this second idea, the procedure is to measure the star magnitudes at the same time as the position is being measured, with this time being presumed to always correspond to the stars being at the same altitude for reasons associated with the positional measure. So for example, suppose that Ptolemy had an observing procedure (perhaps based on some unknown intricacy of the armillary sphere that required constant altitude observations) that always measured the stars when they were at 10° altitude. Then, all stars would be seen at the same altitude and would have no differential extinction.

This second idea has two technical problems. First, the observer will generally not have a reasonable comparison star observed at the same time and same altitude, so they must go on brightness as based on *memory*. Remembered brightnesses are notoriously poor, and the required comparison would lead to a substantially larger than observed scatter. Second, the southerly stars that do not culminate above the observed altitude cannot be measured with this procedure. The observer would have had to do some sort of correction to a higher altitude, or else there would be a significant spike in the m-V plots at the lowest declinations, and this is not seen.

But the biggest problem with this idea is simply that we know that the observers did not use any such procedure that required all their stars to be measured at the same altitude. The *Almagest* gives long details on the use of an armillary sphere to measure positions, and there is no requirement, utility, or any indication that the stars were positioned at some constant altitude. Al Sufi did not measure star positions (he merely precessed those in the *Almagest*) so he certainly had no such procedure. Tycho reports on the details of his methods and even gives his individual measures, so we know that his positioning was dominated by the use of his meridian circle and sextant. The sextant makes observations at any direction in the sky, with the horizon avoided due refraction, while the meridian circle always measures stars over a wide range of altitudes. With this, we know that the second specific procedure was not realized in practice by any of the observers.

In all, it is possible and plausible that some specific procedure could produce automatically extinction corrected magnitudes. The only real possibility for such a specific procedure is something like having the observer await until some pre-calibrated comparison star rises or sets to the same altitude as the near-culmination southerly star. Any such procedure is un-natural, complex, and time consuming, so it would have been undertaken only as an intentional correction for extinction.

### 6.3. *Extinction correction by use of local standard stars from the Almagest*

Star magnitudes must be estimated by comparing the apparent brightness of the star against some standard. One good and plausible set of standards could have been the nearby stars with magnitudes taken from the *Almagest*. For the northerly stars, the star and its comparison would both be at high altitudes and would have closely similar extinction, so the estimated magnitude would have no systematic effect from extinction. For the southerly stars, the star and its nearby comparison would both be at similar altitudes (both low) and would have similar extinctions (both high), so the estimated magnitude would be on the same scale as its nearby star. If the nearby comparison star has its adopted magnitude from the *Almagest*, then *its* magnitude would have already been extinction corrected so the estimated magnitude of the star will also be extinction corrected.

For example, perhaps the star α Cru was adopted as a standard for magnitude 2 as based on it *Almagest* magnitude, and so the estimate of β Cru would naturally be made by a comparison to this. Such a procedure is natural for an observer. This could also account for a small fraction of apparent copying, such as the ~10% estimated for Tycho's catalogue, so it is not copying, but rather simply adopting the earlier work as the standard.

This explanation might work for the catalogues of Al Sufi and Tycho, but it cannot explain the extinction correction in the *Almagest*. As such, this explanation is not satisfying because it merely pushes the question back to another identical question, while offering no answer for how the *Almagest* stars are extinction corrected.

This explanation can be sharply tested, due to the variation in the effective $k$ value over different ranges of RA in the *Almagest*. Recall that the fitted $k$ values are 0.165±0.027 mag/airmass for RA from 135° to 225° and -0.30±0.05 mag/airmass for RA from 270° to 45°. The southern *Almagest* stars would be the local standards, so the nearby stars in later catalogues should show the same fitted $k$ values if this idea is correct. The catalogue of Al Sufi distinctly fails this test, both for the whole catalogue as well as for the subset with greatly different magnitudes (which were certainly not copied). The catalogue of Tycho also distinctly fails this test, both for the original 777 star catalogue and for the full 1004 star catalogue. This is a decisive test, and it shows that the later catalogues are not extinction corrected due to using local standard stars from the *Almagest*.

### 6.4. *Seven failed ideas for extinction correction*

What if the observers were roughly 10° further south in latitude than commonly assumed? Or maybe the cataloguers were using reports from travellers to the south? With this, the extinction correction for the observer would be substantially smaller and might not be noticeable? But Ptolemy was not observing from the Sudan, Al Sufi was not observing from Yemen, and Tycho was not observing from Italy. Reports from travellers are not plausible[18], especially not of quantitative magnitude estimates for many and obscure stars. I strongly conclude that this idea is not consistent with the known history of the cataloguers and of the times.

What if the extinction really was as low as derived? But all measures of pre-industrial sea-level extinction are always with k>0.23 mag/airmass or so. Even if the aerosols were miraculously disappeared (which is impossible because the many natural sources of aerosols are omnipresent), the ordinary extinction from the atmospheric gases is 0.14 mag/airmass. Either of these hard limits is greatly inconsistent with the measured values. I conclude that this idea is unphysical, because it would require the Earth's atmosphere to go away.

Perhaps there is some psychological or physiological effect that makes humans naturally over-estimate the brightnesses of stars near horizon? Suppose that this effect increases towards the horizon with a rough compensation for the atmospheric dimming. Such a hypothetical effect has similarities to the well-known 'Moon Illusion', wherein objects appear larger when viewed near the horizon. But this idea has severe problems with it being unlikely that any such effect would compensate for extinction as exactly as observed. Also, no such effect has ever been recognized to date, and this makes the existence of any such hypothetical phenomenon very unlikely. But the previously-unrecognized existence of such a new phenomenon is already disproven by my data reported in Appendix 4 (see Figure A3).

A similar idea is that perhaps stars appear brighter when seen against a brighter background, and the sky near the horizon is somewhat brighter than higher up[19], so this could make southerly stars appear brighter in compensation for the extinction. This possibility is similar to the case where early star magnitude cataloguers would find that stars in the Milky Way (with surface brightnesses essentially the same as the sky brightness near the horizon[20]) have systematic brightness offsets of 0.06 to 0.25 mag[21]. However, this effect is greatly too small to account for the >1 magnitude extinction effect expected for stars within 10° of the horizon, so we already know that this cannot be the explanation for the extinction correction. Most importantly, the effect goes in the *wrong* direction, with stars in brighter areas being reported *fainter* than their modern magnitudes would indicate. Another simple refutation is to simply point at Figure A3 (in Appendix 4) and say that this effect can only be negligibly small. So this idea is confidently known to be wrong.

A fifth and sixth idea to explain the observed extinction correction is to suppose that the observers used comparison stars that were only slightly higher in the sky than the southerly stars being estimated. With this, the differential extinction effects would be small. The trouble with this idea is that of getting extinction corrected magnitudes for the comparison stars. It might be possible to observe a series of comparison stars at progressively greater zenith distances so as to calibrate southerly comparison stars, where each step has moderately small differential extinction. But this idea fails because the cumulative differential extinction is mathematically identical to making the comparison over just one large step. Or it might be possible that the observers chose their comparison stars to be southerly, in which case the comparisons with the more southerly stars might have only moderate differential extinction. But this possibility fails because then the stars passing overhead will always appear brighter then expected when compared to their southerly comparison stars. That is, within this idea, there must still be a significant difference in m-V for the southerly comparison stars and the more northerly stars, and this effect is clearly absent (see Figures 1-4).

The seventh failed idea (in this subsection) is that perhaps the magnitudes were defined (or measured) by their altitude of first visibility above the horizon. This angle above the horizon of first visibility is called the extinction angle[22]. The idea would be something like first magnitude stars appearing at 'one finger' above the horizon, second magnitude stars appearing at 'two fingers' up, third magnitude stars having an extinction angle of 'four fingers', and so on. With such a scheme, the southern stars would be measured with a standard that is identical to that use for the northern stars. This idea fails because the measured extinction angle is greatly variable, for both of two effects, and this would result in a large scatter in derived magnitudes that is much larger than present in any of the catalogues. One effect is simply that the extinction coefficient changes from day to day and season to season. The typical one-sigma scatter in the aerosol component of $k$ for sites around the world is 40%, while seasonal variations are typically varying by a factor of two in the aerosol component for temperate sites worldwide[23]. For sites close to those of the three observers, this results in daily variations with one-sigma changes of typically 0.10 mag/airmass added on top of seasonal RMS variations are typically 0.10 mag/airmass[24]. This would cause a measurement error in the reported magnitude by roughly one magnitude[25]. The second effect is that the observational measurement error in the extinction angle is large, with this translating into a large scatter in the reported magnitude. I have much experience at observing extinction angles, and I find that the one-sigma scatter in the derived magnitude would be 0.5 mag and 0.8 mag from two sites[26]. If this seventh idea were realized in history, then these two large sources of error would result in RMS scatters in Tables A1-A4 of more than 1.0 mag (with additional scatter coming from quantization errors). This is not seen, so the seventh idea is not realized in history.

# 7. CONCLUSIONS

The basic result of this paper is that the star catalogues of Ptolemy, Al Sufi, and Tycho all report the magnitudes for a thousand stars with the southern stars showing near-zero dimming, which is to say that the magnitudes were extinction corrected. This result is very robust, with the derived $k$ values being highly significantly lower than any physically possible extinction, with this being true for all of the many subsets of the various catalogues, and with the result being apparent by several different methods.

The basic result (that the three star catalogues somehow corrected for extinction) comes as a surprise for historians of astronomy. This is partly due to historians largely ignoring extinction[27], so if no one is thinking about an issue then any new result would be surprising. This surprise is partly because no old source ever talks about atmospheric extinction[28], so the default idea would then be that the old observers did not know about it and hence could not correct for it. The essentially modern theory of extinction was presented by Bouguer[29] in 1729, while de Mairan had a faulty and vague description in 1723, but I know of no prior explanation (scientific or otherwise), and I even know of no prior recognition of the existence of the extinction phenomenon[30].

The central question is *how* these widely spread out magnitudes came to be extinction corrected? I have discussed many possibilities, each with variants. Some of these ideas are definitely against strong evidence or impossible. I have discussed these ideas in detail because colleagues have repeatedly been suggesting these ideas to me, so they are alluring and some researchers might settle on them without due consideration. Here, I will list these eleven wrong ideas and the reasons why we know that they are wrong with high confidence: (1) Perhaps the low apparent extinction is just some sort of a horrible mistake or fluctuation in the data? This idea is strongly refuted by the very high statistical significance of the fitted *k* being greatly lower than any possible extinction, as well as by the robustness of the result being significant for all three catalogues, for stars over various ranges of southerly declinations and right ascensions, for various subsets of the catalogues, and for several methods of measure. (2) Perhaps the observers used some modern-style calibrated mathematical model to remove the effects of extinction? This idea is an anachronism, where the old observers did not have the required knowledge. (3) Another wrong idea is that the observer was estimating the magnitudes at the same time as they measured the positions and that there was some procedural reason for observing the stars all at the same altitude, thus making for near-zero differential extinction. This idea fails because we know the procedures for measuring positions for all three observers and none of them used constant altitudes. (4) Al Sufi and Tycho might have used the *Almagest* stars far to the south as local standard stars, so their resulting southern magnitudes would be extinction corrected simply because the *Almagest* was already extinction corrected. This idea is refuted by the utter lack of the local anomalies that appear in the *Almagest* being duplicated in the later catalogues. (5) Perhaps the observer looked from some site over 10° in latitude further south than their known observatories, or perhaps they used extensive quantitative reports from travelers to the far south that gave magnitudes? But all such ideas are unhistorical because we know where the observers observed from and because travelers reports of hundreds of faint stars is not plausible. (6) Or perhaps the extinction really was as low as given by the fitted *k* values? But this idea is unphysical because the fitted values would require an Earth with no atmosphere.  (7) Perhaps there is some illusion, like the famous Moon Illusion, whereby stars near the horizon are perceived as being brighter than a photometer would measure?  But such a hypothetical phenomenon has never been reported, and indeed is already disproven by the data shown in Figure A3.  (8) Perhaps the factor-of-two brightening of the sky towards the horizon results in the stars being perceived as being brighter than a photometer would measure, with this perhaps compensating for the atmospheric dimming?  This idea fails because the effect actually works in the wrong direction, it makes stars appear *fainter*, and because this phenomenon must be negligibly small as shown in Figure A3.  (9) Perhaps the comparison stars for the southerly stars were only slightly more northerly so that differential extinction is small?  But to calibrate the comparison stars requires that they be connected to the overhead stars by one or more steps, each with further differential extinction, and mathematically the use of one or many steps is identical to simply comparing the southerly stars to overhead stars, so this suggestion is not a solution.  (10) Perhaps the observers chose all their primary standard stars to be in the south so that then there would be no differential extinction between them and the other southerly stars?  But this is refuted by the more northerly stars not appearing brighter than the standards.  (11) The last failed idea is that perhaps the magnitude system was defined or measured by the altitude of first visibility of stars, and this use of extinction would produce the same calibration for all declinations.  This idea fails because the real variations in observing the altitude of first visibility lead to a much larger scatter in magnitudes than is observed for any magnitude range of any of the catalogues.

These eleven wrong ideas can only cause confusion for the good ideas as to how the reported magnitudes are given with extinction corrections.  There are a variety of ideas that are reasonable explanations, and I do not know how to decisively select only one of these as being the answer.  Further, there are variants on each of these ideas that are plausible, and the historical reality could easily involve more than one of these reasonable ideas in combination.  Here are the three basic possible and plausible ideas that can readily explain the extinction corrections in the three star catalogues:

**(A)** The existence of extinction as a phenomenon is readily recognizable by any thoughtful observer, especially if they are paying attention to star brightnesses.  So it is plausible that the observers knew about extinction on a purely empirical basis.  All three old observers were inside a culture that strived for accuracy and frequently made quantitative corrections to measures, so it would have been natural for them to correct for the dimming of extinction.  Such corrections might have been as simple as the observer intentionally-or-unintentionally changing their observed brightness to a brighter magnitude bin, especially if the observed brightness was judged to be near the edge of the observed bin, as calibrated by their personal observations of the dimming of setting stars.  Such empirical corrections do not require any knowledge, sophistication, or equipment on the part of the observer.  Indeed, humans are frequently making such corrections on an unconscious basis, so the observers need not even have recognized the existence of the extinction phenomenon.

**(B)** As a middle ground between ideas 2 and A, above, it is possible that the observers had a more formal mechanism, still purely empirical, for correcting the observed magnitudes into the reported magnitudes. We know that Tycho observed the setting of bright stars so as to construct refraction tables, so it is plausible that the observer watched the dimming of setting stars (or the brightening of rising stars) so as to construct a table of extinction corrections as a function of altitude. This table (or graph or set of rules-of-thumb) would then be applied to observed magnitudes of the stars (given their altitudes at the times of observation) so as to get the reported magnitudes. This idea requires no sophistication, equipment, or modern knowledge. This idea is proven to be practical, easy, fast, and effective by my modern application, as shown in Appendix 4.

**(C)** The third possible and plausible means by which the three star catalogues could produce extinction corrected magnitudes is if the observer intentionally observed all the stars at some constant altitude. That is, by seeing the target star and the comparison star at the same altitude (say, 10° above the horizon), the extinction to both will be identical, so the southerly stars will have little differential dimming due to extinction. Idea C is possible and plausible, but the mechanics of the observations are awkward, slow and with difficulties. One difficulty is that the observer will need some means, perhaps crude, for measuring the altitude of stars so as to know when the comparison star has the same altitude as the target. Another difficulty is that the window in time for making the observation is small, roughly a dozen minutes for the southerly stars, and the observer must wait for this condition to occur. Another difficulty is that the southern stars can only be estimated after a network of perhaps 50 northern stars spread all around the sky has been previously calibrated. All of these difficulties can be overcome by a willful observer, but at the cost of a lot of time and trouble. Idea C is so awkward and contrived that it would arise only if the observer knew about extinction and decided on this procedural method to correct the magnitudes.

So which is it, idea A, B, or C? If you think that the old observers did not know about extinction, then you would say that idea A is correct, with the offsets being applied on a purely instinctive basis. If you think that the old observers did know about extinction, than any of ideas A, B, and C are reasonable to choose from. If you think "ancient astronomers are smarter than we knew"[31], then ideas B and C would be appealing. If you think that the observers did not have sophistication, as there were no giants to stand on the shoulders of, then idea A is the choice. I know of no decisive argument to choose between the possibilities, so I would be happy with any of A, B, or C. If I had to select only one answer, then I would quickly choose B, because the existence of extinction as an empirical phenomenon is easy to recognize, because the culture of all three observers was to make quantitative corrections from observation-based tables and because idea C is much more awkward and time consuming than idea B.

13. Jean-Jacques de Mairan, "Eclaircissement sur le memoire de la cause generale du froid en hiver, & de la chaleur en eté.", *Mem. Acad. Roy. Sci.* (Paris, 1723), 8-17.
14. The recognition of extinction as a phenomenon is easy to see by any observer, because stars dim by many magnitudes as they get close to the horizon.  As such, it is inevitable that many people (from prehistoric times till the Renaissance) independently came to an empirical realization that atmospheric extinction exists (i.e., that stars appear dim as they get near the horizon, for whatever reason).  So it is surprising that there are no surviving records that recognize extinction as a phenomenon.  I have very broad knowledge on similar topics in astronomical history, and I have consulted a number of astronomy historians with very broad and deep knowledge (including Gary Thompson, Andrew Young, Owen Gingerich, as well as the many readers of the HASTRO-L group), with null results.  While it is effectively impossible to prove a negative (that no pre-1723 sources mention extinction), the lack of any old sources does suggest that formal knowledge of extinction was not widespread.
15. Schaefer, "Latitude" (ref. 4).
16. The early catalogues of magnitudes by William Herschel, John Herschel, and Friedrich W. A. Argelander did not have extinction corrections, but care was taken to avoid to avoid low altitudes.  The first large scale photometry catalogues to explicitly correct for extinction was in 1872 by Eduard Heis for his *Atlas coeletis nova*, where he consciously corrected his observed magnitudes for extinction.  In 1872, Benjamin Gould made observations for the *Uranometria Argentina* by setting up secondary standard stars by making looking at them while the northern primary standards were at the same altitude.
17. To use an everyday example, humans are quite good at estimating the height of people seen in the distance, where we unconsciously know to scale the estimated height (from the observed angular height) by the inverse of the estimated distance.  This happens despite most humans having no idea of the existence of what astronomers call the 'small-angle formula'.  In a closer example, involving naked eye photometry, we have the everyday task of a driver on a highway at night viewing streetlights near the horizon.  All humans naturally and intuitively will take the observed brightness and correct for the distance so as to estimate the luminosity and type of the light, even though most humans have never heard of the inverse-square-law for light.  The point is that human are all the time making such corrections to their perceived measurements, and this is often done unconsciously, so it is easy to think that the old observers could have made an intuitive correction for the dimming of stars near the horizon.
18. The use of a traveler's report is plausible only for singular cases involving some famous star.  The only possibilities for this are Achernar at the end of Eridanus and Canopus.  Canopus (α Car) had a declination of -52.5° in the time of

Ptolemy, so it would culminate with an altitude of 6.3° above the southern horizon of Alexandria. Canopus has a negative magnitude of -0.70, so with k=0.25 mag/airmass, it will always appear dimmed by at least 1.88 mag, so it would never appear brighter than m=1.18. For Ptolemy, it would be appropriate still to call it a first magnitude star as based on direct observation without correction. From my university campus in Baton Rouge Louisiana ($\lambda$=30.4°), Canopus culminates 6.9° above the horizon, and I can always astound students by pointing to it and having them compare it with Rigel, even on the clearest of winter evenings. For Hipparchus as observing from Rhodes ($\lambda$=36.4°), the culmination is at 0.9° above the southern horizon, and it actually starts to matter that the refraction correction raises this to 1.4°. For the best plausible extinction for sea level for an eastern Mediterranean site (0.23 mag/airmass), the dimming (with a zenith distance of 88.6° and X=23 airmass) is 5.05 mag at culmination, so Canopus will never appear brighter m=4.35 mag. Thus, it is that Canopus is certainly never a first magnitude star as seen by Hipparchus. Nevertheless, this is not a strong argument against a Hipparchan source for the *Almagest* magnitudes, because the magnitudes are already extinction corrected. Or maybe Hipparchus included Canopus simply due to travellers reports.
19. My simple physical model for the brightness of the dark nighttime sky (K. Krisciunas & B. E. Schaefer, "A model of the brightness of moonlight", *Publications of the Astronomical Society of the Pacific*, ciii (1991), 1033-9, see equations 2 and 3) gives the brightness 10° above the horizon to be 2.0 times the brightness near the zenith. A more detailed physical model (R. H. Garstang, "Night-sky brightness at observatories and sites", *Publications of the Astronomical Society of the Pacific*, ci (1989), 306-29) shows that the sky brightens from the zenith to 10° altitude by close to a factor of two, but then greatly darkens going to 0° altitude. Observations (E. O. Hulburt, "Night sky brightness measurements in latitudes below 45°", *Journal of the Optical Society of America*, xxxix (1949), 211-15) show the dark and clear sky brightness peaks at around 10°-15° altitude with the peak being 50%-70% brighter than the zenith.
20. The Milky Way region comes to a peak brightness close to a factor of 2 times the zenith sky brightness at high galactic latitudes. See Hulburt "Night sky" (ref. 19), and C. W. Allen, *Astrophysical quantities* (London, 1973, 3rd ed.), p. 134. This means that the systematic biases in estimating magnitudes for stars in the Milky Way are essentially the same as for measuring stars near the horizon.
21. Lundmark "Luminosities" (ref. 1), p. 250, 254, and 256.
22. B. E. Schaefer, "Atmospheric extinction effects on stellar alignments", *Journal for the history of astronomy, Archaeoastronomy supplement*, xvii (1986), S32 gives a full physics model plus many observations of extinction angles.

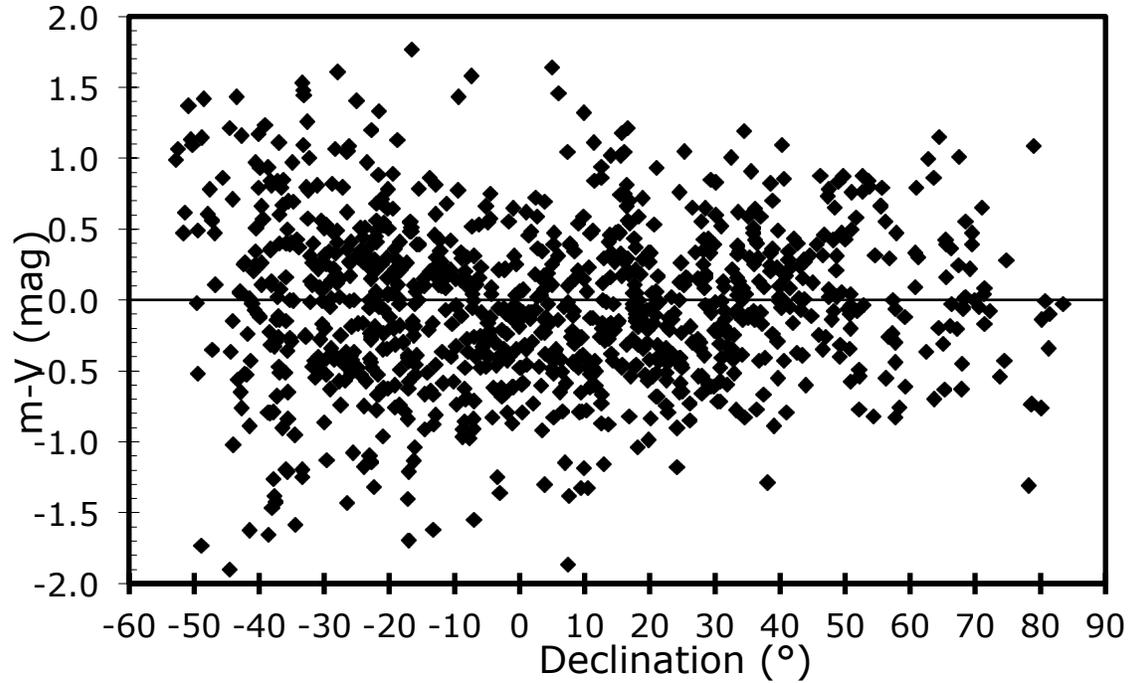

**Fig. 1. Dimming of stars for the *Almagest*.**
The m-V value for each star is the deviation of the reported magnitude in the *Almagest* as compared to the modern magnitude, and this depends on the atmospheric extinction, the usual observational uncertainties, and the quantization of magnitudes into bins. The effect of atmospheric extinction is to make the southern stars appear dim (positive m-V), so the points should show a sharp rise to the left in this plot (just as seen in Figure A2 and A3). Startlingly, this extinction dimming (the upturn to the left) is *not* seen. That is, the southern stars, that can only be seen low above the southern horizon, are reported to have magnitudes just as bright as if there were no atmospheric dimming. So, the magnitudes reported in the *Almagest* are already corrected for extinction. A formal chi-square fit to an extinction model (equation A4) gives an extinction coefficient of +0.010±0.017 mag/airmass, with this being greatly smaller than the smallest possible for a pre-industrial sea-level temperate-latitude site (~0.23 mag/airmass). From this plot, we can visually see the main result of this paper, that the ancient star catalogue reports magnitudes are fully extinction corrected.

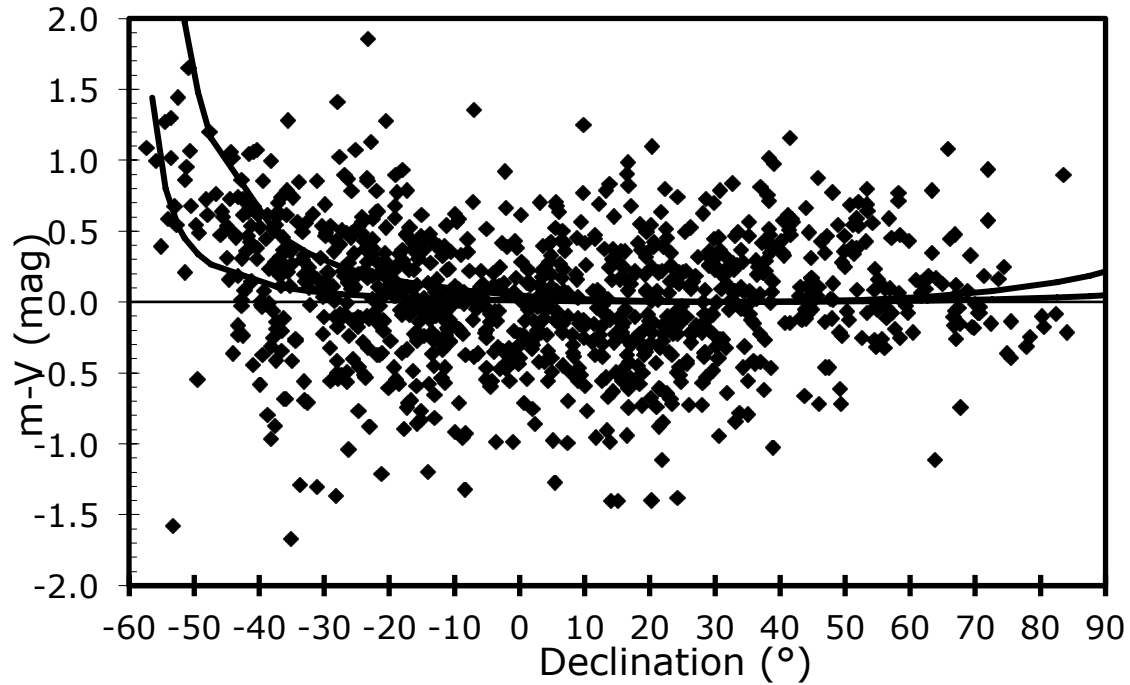

**Fig. 2. Dimming of stars for Al Sufi's catalogue.**
The dimming of Al Sufi's reported magnitudes (m-V) shows a small upturn to the left. The best fit extinction coefficient is +0.057±0.007 mag/airmass, with the model shown as the lower curved line. This extinction is greatly and significantly less than the expected minimal value of 0.25 mag/airmass (with the model shown by the upper curve), and this can be seen easily in this plot because almost all the southern stars are far below the model curve. The fitted value shows that some form of extinction correction was applied, with this only resulting in ~75% of the extinction being corrected for. Thus, the catalogue of Al Sufi is reporting magnitudes that are extinction corrected, even thought this correction is imperfect with a 25% error.

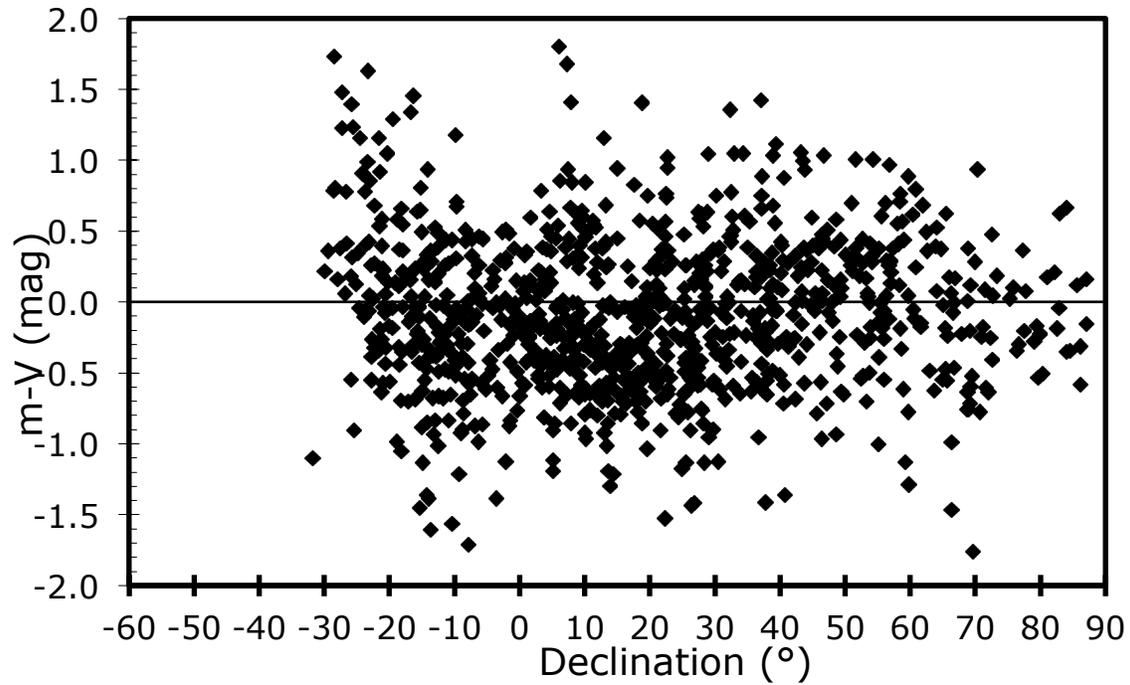

**Fig. 3. Dimming of stars for Tycho's 1004-star catalogue.**
The magnitudes reported for the southern stars in Tycho's 1004-star catalogue do not have a sharp upturn to the left as expected from dimming caused by atmospheric extinction. The best fit extinction coefficient to the plotted data is +0.013±0.012, which is consistent with zero extinction. This is to say that Tycho's magnitudes are reported to us with an extinction correction.

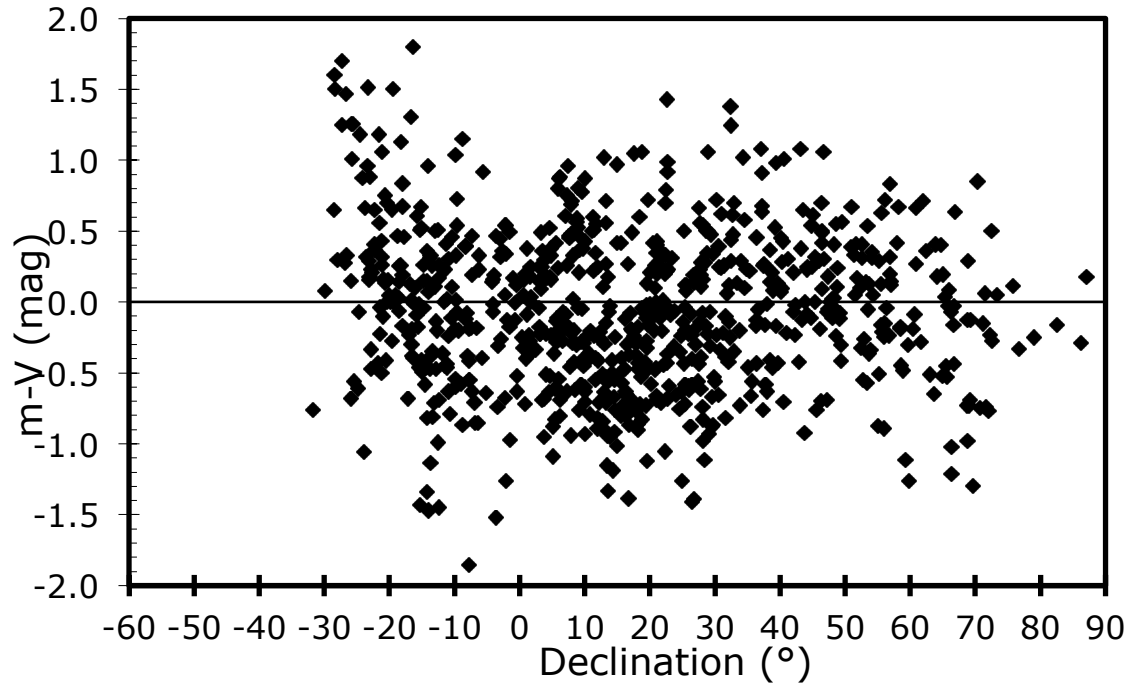

**Fig. 4. Dimming of stars for Tycho's 777-star catalogue.**
Again, there is no obvious upturn in the dimming for the reported magnitudes as the southern limit is approached. The best fit extinction coefficient is +0.044±0.011 mag/airmass, significantly and substantially below the lowest possible value (~0.23 mag/airmass) for a sea-level pre-industrial site. This is to say that Tycho's original 777-star catalogue reports magnitudes that are extinction corrected. There is a small excess of ~15 stars with m-V greater than 1.0 for declinations south of -17°, with all these stars being in the Sagittarius/Capricorn area. To create this localized anomaly, Tycho's extinction correction procedure, whatever it was, only provided half the correction to zero extinction.

# APPENDIX 1:  CONVERSION TO MODERN MAGNITUDES

The modern magnitude system has its origin in the ancient Greek magnitude system. Now, magnitudes are quantified by a logarithmic equation,

$$m = m_0 - 2.5 \, \text{Log}_{10}(F/F_0), \tag{A1}$$

where 'F' represents the measured flux above the atmosphere and the '0' subscript refers to the values for some standard star.  Subtle differences in the definition for the magnitude scale can lead to small (~0.02 mag) changes in the zero of the magnitude scale.  In pre-modern times, the magnitude system was apparently viewed only as a way of grouping stars by brightness.  For example, stars labeled as 'second magnitude' all had similar brightnesses and were perceived as being brighter than 'third magnitude' stars and fainter than 'first magnitude' stars.  Both Ptolemy and Al Sufi subdivided each magnitude group into additional subsets that were brighter and fainter than the group average, effectively reporting the star magnitudes with a precision of roughly one-third of a magnitude.

By construction, the modern logarithmic scale follows along the lines of the ancient scale, nevertheless, there are substantial deviations.  The largest systematic deviation is that the fainter stars are reported to be substantially brighter than they are on the modern scale.  For example, the stars labeled as 'sixth magnitude' in the Almagest have an average modern magnitude of 5.20.  So a report that a particular star is 'sixth magnitude' should not be interpreted as saying 'm=6', but rather as saying that the star is in a group whose average magnitude is 5.20.  The connection between any old magnitude bin and the corresponding modern magnitude can only be established empirically.

The connection between the old and modern magnitudes can be measured from the distribution of the modern V-band magnitudes for all stars within each catalogued magnitude group.  For example, the Almagest reports on 49 stars as being 'sixth magnitude', and these stars have an average modern magnitude of 5.20 with an RMS scatter of 0.37 mag.  That is, whenever Ptolemy says a star is 'sixth magnitude', we should translate this as the star having been reported as V=5.20±0.37.  The RMS range corresponds to the 1-sigma range over which roughly 68% of the stars are within, thus close to two-thirds of Ptolemy's 'sixth magnitude' stars have 4.83<V<5.57, while the rest of the stars are closely outside this range.  This small scatter is actually encouraging because it implies a good photometric accuracy.  This accuracy is comparable to the typical uncertainty in human photometric accuracy under non-optimal conditions and is comparable to the quantization error introduced by reporting magnitudes to a precision of only something like a third of a magnitude.  The typical RMS scatter for other magnitudes and the other catalogues is more like half a magnitude, which is still usable and good accuracy.

The main point of this paper concerns the effects of atmospheric extinction, which are minimal for stars with culminations in the sky of less than 1.1 airmass.  (For a good night extinction coefficient of 0.25 mag/airmass, the difference in apparent brightness as a star moves from the zenith to 25° away is only 0.025 mag, with this being imperceptibly small.) That is, for stars that pass within 25° of zenith, we can expect that extinction effects will be negligible. For the 'sixth magnitude' stars in the Almagest that have declinations from +12° to +55° (so as to allow for the disputed latitude of the observer) have an average of m=5.27. This would be the relevant value for calibrating all 'sixth magnitude' stars if we need to make certain that the calibration is unaffected by any extinction problems.

For each of the pre-telescopic catalogues (including the two version of Tycho's star catalogue) and each of the magnitude bins, I have calculated the average modern magnitudes for all stars and for all stars that culminate within 25° of the local zenith. These results are tabulated in Tables A1 through A4. The columns present the magnitude bin label in the old catalogue, the total number of stars reported in that magnitude bin, the average modern magnitude for these stars ($<m>$), the RMS 1-sigma scatter of these magnitudes, and the average modern magnitude for stars that culminate within 25° of local zenith ($<m>_z$). The last column serves as the best estimate (i.e., with no questions involving extinction) of the perceived modern magnitude for a star with a given catalogue label. To give another example, when Al Sufi reports a star as having magnitude '3', we should interpret this as a measure of the apparent brightness of the star being m=2.91±0.56 mag.

## TABLE A1, Modern magnitudes for Ptolemy

| Ptolemy Magnitude | Number Stars | $<m>$ | RMS | $<m>_z$ |
|---|---|---|---|---|
| 1 | 13 | 0.42 | 1.07 | 0.37 |
| <1 | 2 | 1.32 | 1.16 | 2.14 |
| >2 | 3 | 1.55 | 0.73 | 2.23 |
| 2 | 35 | 2.11 | 0.83 | 1.74 |
| <2 | 7 | 2.85 | 0.69 | 2.35 |
| >3 | 12 | 2.50 | 0.47 | 2.41 |
| 3 | 174 | 3.17 | 0.67 | 3.11 |
| <3 | 21 | 3.74 | 0.82 | 3.25 |
| >4 | 84 | 3.84 | 0.57 | 3.81 |
| 4 | 371 | 4.26 | 0.57 | 4.22 |
| <4 | 18 | 4.74 | 0.60 | 4.92 |
| >5 | 9 | 4.70 | 0.59 | 4.70 |
| 5 | 206 | 4.82 | 0.50 | 4.84 |
| 6 | 49 | 5.20 | 0.37 | 5.27 |
| faint | 12 | 4.57 | 0.35 | 4.57 |

**TABLE A2, Modern magnitudes for Al Sufi**

| Al Sufi Magnitude | Number Stars | $<m>$ | RMS | $<m>_z$ |
|---|---|---|---|---|
| 1 | 13 | 0.51 | 1.16 | 0.74 |
| 1-2 | 2 | 0.74 | 0.34 | 1.00 |
| 2-1 | 2 | 0.69 | 0.11 | 1.00 |
| 2 | 26 | 1.78 | 0.44 | 1.84 |
| 2-3 | 7 | 2.32 | 0.28 | 2.32 |
| 3-2 | 11 | 2.33 | 0.34 | 2.41 |
| 3 | 95 | 2.80 | 0.56 | 2.91 |
| 3-4 | 95 | 3.44 | 0.46 | 3.48 |
| 4-3 | 75 | 3.67 | 0.44 | 3.67 |
| 4 | 264 | 4.16 | 0.51 | 4.15 |
| 4-5 | 86 | 4.50 | 0.49 | 4.61 |
| 5-4 | 22 | 4.57 | 0.44 | 4.64 |
| 5 | 167 | 4.65 | 0.46 | 4.72 |
| 5-6 | 67 | 5.02 | 0.45 | 5.12 |
| 6-5 | 1 | 4.69 | ... | 4.69 |
| 6 | 70 | 5.15 | 0.37 | 5.13 |
| 6-7 | 8 | 5.55 | 0.34 | 5.77 |

**TABLE A3, Modern magnitudes for Tycho's 1044-star catalogue**

| Tycho Magnitude | Number Stars | $<m>$ | RMS | $<m>_z$ |
|---|---|---|---|---|
| 1 | 12 | 0.69 | 0.98 | 0.06 |
| 2 | 41 | 2.07 | 0.66 | 2.18 |
| 3 | 163 | 3.35 | 0.69 | 3.24 |
| 4 | 329 | 4.16 | 0.57 | 4.04 |
| 5 | 227 | 4.64 | 0.47 | 4.65 |
| 6 | 198 | 5.07 | 0.42 | 4.92 |

**TABLE A4, Modern magnitudes for Tycho's 777-star catalogue**

| Tycho Magnitude | Number Stars | $\langle m \rangle$ | RMS | $\langle m \rangle_z$ |
|---|---|---|---|---|
| 1   | 8   | 0.40 | 0.91 | 0.40 |
| 1 . | 4   | 1.27 | 0.96 | 1.27 |
| 2 : | 5   | 1.03 | 0.75 | 1.25 |
| 2   | 29  | 2.21 | 0.46 | 2.19 |
| 2 . | 7   | 2.19 | 0.70 | 2.44 |
| 3 : | 13  | 2.88 | 0.55 | 2.28 |
| 3   | 101 | 3.25 | 0.62 | 3.10 |
| 3 . | 33  | 3.62 | 0.78 | 3.71 |
| 4 : | 22  | 3.68 | 0.62 | 3.70 |
| 4   | 188 | 4.12 | 0.60 | 4.07 |
| 4 . | 53  | 4.21 | 0.45 | 4.18 |
| 5 : | 26  | 4.48 | 0.50 | 4.53 |
| 5   | 122 | 4.57 | 0.45 | 4.62 |
| 5 . | 27  | 4.74 | 0.58 | 4.87 |
| 6 : | 11  | 5.14 | 0.32 | 4.85 |
| 6   | 100 | 5.16 | 0.37 | 4.83 |
| 6 . | 20  | 5.08 | 0.49 | 5.61 |

# APPENDIX 2: EXTINCTION MODEL

Extinction is the dimming of starlight as it passes through the Earth's atmosphere. This depends on the product of two quantities that vary from observation to observation. The first quantity is the airmass, X, which is the optical pathlength through the atmosphere. This depends primarily on the zenith angle, Z, which is the angle between the zenith and the star. For a sightline pointing exactly to the zenith, with Z=0°, the airmass is unity, while for a star that appears 60° from zenith, the airmass is X=2.00. The second quantity is the extinction coefficient, $k$, which is the magnitude dimming for light passing through one airmass of atmosphere. The dimming of the starlight will be by a magnitude equal to kX and a factor equal to $10^{-0.4kX}$.

The airmass for a flat Earth would be simply X=sec(Z). This is an excellent approximation for all observations far away from the horizon. As the horizon is approached, the airmass goes to roughly X=40 (instead of the infinite value as taken from the flat Earth model), with the exact value depending on the vertical distribution and density of the aerosols. The total airmass is

$$X=[\cos(Z) + 0.025 \, e^{-11\cos(Z)}]^{-1}, \quad (A2)$$

to an excellent approximation[32].

The extinction coefficient, $k$, is for the visual bandpass, and this varies widely from site to site and night to night. The extinction has strong correlations with the elevation above sea level, the month of the year, the relative humidity, and the latitude[33]. Nevertheless, there are substantial variations even after these effects are accounted for. We do not know the conditions for any particular observations in any of the star catalogues, so all we can do is note site averages and recognize the level of variations for each site. For the Almagest star catalogue, I have previously given an exhaustive analysis on the effective value of $k$ as based on data from many sources[34]. The conclusion is "The plausible range of k is from 0.23 to ~0.4 mag. The most likely value is probably 0.25 mag, as taken from the heliacal rise data of Ptolemy". Further, under no plausible conditions will $k$ be smaller than 0.23 mag or so, even with selection of good nights by the observer. I find similar conclusions for the observing sites of Tycho and Al Sufi.

Astronomical magnitudes are compared against a standard star with the fluxes as measured above the atmosphere, which is to say that all observed fluxes are corrected to zero atmosphere. These classical astronomical magnitudes (as tabulated in tables and catalogues in the modern literature) are denoted with the symbol $V$, so equation A1 gives $V = -2.5 \, \text{Log}_{10}(F/F_0)$, where $F$ and $F_0$ are the fluxes of the star and the zero-magnitude standard, respectively, above the atmosphere. But for visual observations, magnitudes are measured with respect to other stars as seen from the bottom of the atmosphere. The observed flux for a star at airmass $X$ will be $10^{-0.4kX}F$, while the observed flux for the zero-magnitude standard at zenith is $10^{-0.4k}F_0$. The reported magnitude will always be made with respect to a comparison star, and in general this comparison star will be high in the sky. With equation A1, the apparent magnitude of a star will be

$$m = V + k(X-1). \tag{A3}$$

When combining equations A2 and A3, we now have a complete model for how bright a star should be reported by any observer.

When a star is observed around the zenith, the observed brightness will be close to V, as tabulated in modern star catalogues. This result is a good approximation for a large area of the sky away from the horizon. For example, for stars within 25° of the zenith (Z<25°), we have X<1.10, and the error (m-V) is <k/10, which for k=0.25 mag/airmass is 0.025 mag, which is so small as to be not observable. For stars at Z=60° (so X=2.00) and k=0.25 mag/airmass, we expect systematic discrepancies between *m* and *V* of a quarter of a magnitude, and this is observable in the statistics of many stars. For stars 5° above the horizon (Z=85°, X=10.3), the star appears 2.3 mag fainter than if it was at zenith. For the extreme case of a star on the horizon (Z=90° so X=40 or so) and k=0.25 mag/airmass, the observed magnitude is about 10 mag fainter than V, which is to say that stars at the horizon are always too faint to be visible.

The natural observing conditions is for the star to be somewhere near culmination. For the comparison stars that pass directly overhead, the airmass is <1.10 for stars seen anytime within a time interval of 7.0 hours centered on culmination. For stars that culminate 10° above the southern horizon, the airmass varies by less than 0.1 from culmination over a 100 minute time interval centered on culmination. (I have made extensive observations of stars near the southern horizon, and I confirm that star brightness does not change noticeably over a long interval centered around culmination.) To a reasonable approximation, the time of observation will have the same extinction as at culmination, at which time the zenith distance is given as Z=|λ−δ|, where λ is the observer's latitude and δ is the star's declination. Now we can give the full model for the effects of extinction as a function of the declination of the star:

$$m-V = k([\cos(\lambda-\delta) + 0.025\, e^{-11\cos(\lambda-\delta)}]^{-1} - 1) \tag{A4}$$

Figure A1 illustrates the behavior of m-V as a function of δ for the default case of the Almagest (k=0.25 mag/airmass and λ=31.2°). For comparison, Figure A1 also plots the case where each of the two variables change one at a time, with one curve for the median extinction (k=0.4 mag/airmass) and the other for Hipparchus' latitude (λ=36.4°).

This model can be traced out by the reported stellar magnitudes, so if there was only small error in the reported magnitudes then the plot of m-V versus the star's declination should be close to curves as in Figure A1. However, the real magnitude measures have substantial scatter, both due to the quantization of the reported magnitude values as well as due to the usual measurement errors. From Appendix 1, we see that the typical RMS scatter in m-V (for stars that pass high overhead) is half a magnitude, so this is the sort of scatter that we expect. To display the observations, I have adopted a standard plot showing the m-V for each star as a function of the star's declination. We can see the expected scatter for the many stars, and with the many stars our eyes can readily detect an upturn in m-V towards southerly declinations. This same plot can also be used to display model predictions (like in Figure A1). In this paper, the plots will have identical declination ranges (-60° to +90°) and m-V ranges (-2 to +2 mag) so as to make comparisons easy.

To educate our eyes when looking at such a plot, and to illustrate the model, I have created a simulated set of magnitudes that might be expected for Ptolemy as an observer (with $\lambda$=31.2° and k=0.25 mag/airmass, see Figure A1). For each star, I have added the model m-V to random Gaussian noise with a standard deviation of 0.5 mag for each star. Here, I have used the declinations for all thousand stars in the *Almagest* catalogue. The resultant simulated catalogue of magnitudes is then plotted in Figure A2. As expected, we see a large rise in m-V towards southerly declination, with all stars south of -45° declination having positive m-V. The many stars allow for the best fit curve to be fairly well-defined, and the average value for m-V is significantly positive south of a declination of roughly -25°.

For any catalogue of star magnitudes and declinations (real or simulated), we can fit them to the model in Equation A4 by means of a standard chi-square fit. In these fits, the m-V of all the stars is compared to the m-V as predicted by the model. This difference in m-V (observed minus model) is compared to the uncertainty, $\sigma$, which will equal to the average RMS scatter for stars culminating high overhead (as tabulated in Appendix 1). This ratio is then squared. A star that happens to agree closely with the model will have this ratio near zero, while an average deviation will have the ratio near unity, and a discrepant point will have a ratio substantially larger than unity. The chi-square value ($\chi^2$) is the simple sum of these squared-ratios for all the stars. Importantly, the $\chi^2$ will be the smallest for the best fit, so the procedure is to vary both $\lambda$ and *k* until $\chi^2$ is minimized so as to find the best fit latitude and extinction. Also importantly, the formal one-sigma (68%) error bars can be quantitatively calculated as that range of parameters ($\lambda$ and k) over which the $\chi^2$ is within 1.0 of the minimum value, while the three-sigma (99.73%) confidence range is for which the $\chi^2$ is within 9.0 of the minimum value. The model has two free parameters, the latitude and the extinction coefficient. The number of degrees of freedom in this fit equals to the number of stars minus the number of free parameters, which is always close to 1000. This chi-square formalism has great advantages over the many *ad hoc* statistical criteria that we can invent; with the chi-square methods being very well known, standard, and widely understood. In addition, the chi-square method provides quantitative error bars. As such, chi-square fits are perfect for our analysis of the old magnitudes.

A formal requirement for the chi-square analysis is that the error distribution (in m-V) should be Gaussian. Actually, even substantially non-Gaussian distributions will produce virtually identical best fit parameters as well as reasonable error estimates for these parameters. Nevertheless, the distribution of m-V is closely Gaussian. This can be readily seen by looking at Figures 1-3.  Quantitatively, for example for the *Almagest* values, the distribution is a good Gaussian, where a fit to the observed distribution with a Gaussian returns an average of -0.04 mag and a chi-square of 56 for 62 degrees of freedom.  Alternatively, with a Kolmogorov-Smirnov test between the observed cumulative distribution versus the integral of a Gaussian distribution, the maximum deviation is 0.030 (which is already so small as to point to a good Gaussian distribution), which for N=1016 gives a probability of 0.32 for the two distributions being different, and this says that the *Almagest* m-V values are a good Gaussian distribution.  To further illustrate this, the *Almagest* m-V distribution has 53±7 stars that deviate by more than 2-σ, whereas 46 are expected from a Gaussian, with the point that the *Almagest* does not have any excessive number of outliers.  The same holds true for the other catalogues.  Thus, we can use Gaussian statistics with confidence.

      To illustrate the chi-square fits, I have fit the simulated magnitude catalogue in Figure A2.  With this, I derive the best fit latitude of 31.28° and a best fit extinction of 0.255 mag/airmass.  This is close to the input values, with the reason for this closeness being that we have several hundred stars south of -25° declination.  The chi-square for this fit is 1002.01 (for 1014 degrees of freedom), so the one-sigma error bars are the range of parameters for which the resultant $\chi^2$ is smaller than 1003.01.  With this, the one sigma error bar in latitude is ±0.56°, while the one-sigma error bar in extinction is 0.014 mag/airmass.  This proves that this simple and standard analysis method can determine the average observing latitude to within a fraction of a degree for realistic data.  That is, at this late epoch, we could distinguish the north-south position of the observer to better than 60 miles.  A problem with this is that the latitude and extinction can trade off against each other, for example as seen in Figure A1 where the model for Hipparchus (36.4° latitude and 0.25 mag/airmass) gives closely the same curve as for Ptolemy under an average sky (31.2° latitude and 0.4 mag/airmass).  This trade off will be one-sided because the extinction value is already near the real minimum, so the limit will be stiff against moving north, while a fairly far southerly latitude can be produced with a high extinction.  For the case of the simulated magnitude catalogue, the latitude of Hipparchus ($\lambda$=36.4°) can be rejected both because the best chi-square equals 1029.49 (and is rejected at over the five-sigma confidence level) and because the implied extinction is the unphysical k=0.13 mag/airmass.  This chi-square fit to the simulated data shows that if an astronomer received these data, then they would derive the correct latitude to within roughly 40 miles.

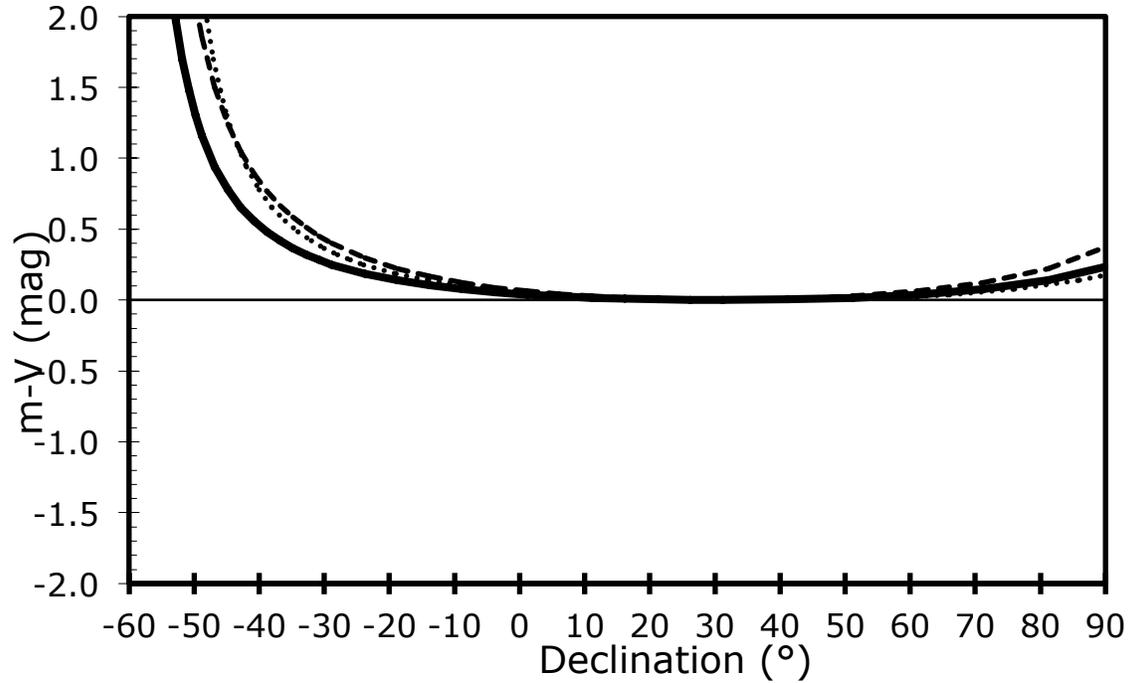

**Fig. A1. Dependence on declination, extinction, and latitude.**
Stars passing high overhead are not dimmed significantly, while stars far to the south can only be seen through a lot of atmosphere and hence will always appear dimmed substantially. Equation A4 presents a model for the amount of this dimming (m-V in magnitudes) as a function of the star's declination, the site's latitude, and the haziness of the air (quantified by the extinction coefficient k). The thick curve shows the model prediction for the case of Alexandria ($\lambda$=31.2°) with the extinction coefficient of k=0.25 mag/airmass. This is the expected case for Ptolemy being the observer of the magnitudes in the *Almagest* star catalogue. This extinction coefficient is for close to the best possible atmospheric clarity, so the dashed curve shows the case for k=0.4 mag/airmass. For the southern stars within ~30° of the horizon, substantial differences will be reported. For the case that Hipparchus was the observer for the *Almagest* star catalogue ($\lambda$=36.4°) with near the minimal atmospheric dustiness (k=0.25 mag/airmass), the model gives the dotted curve. If the reported magnitudes come to us with corrections for these extinction effects, then the observations should follow along the flat line at m-V=0. With this, we see that (if there has been no correction for extinction) we can readily determine the latitude and extinction for the observations.

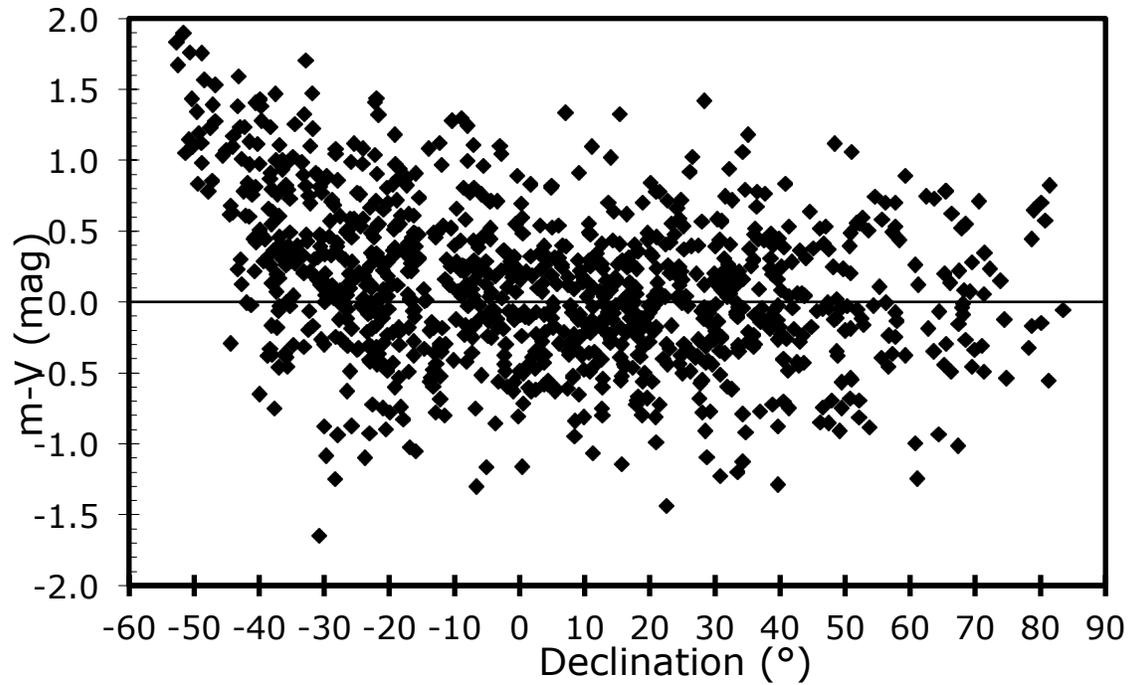

**Fig. A2. Simulated magnitude catalogue.**
I have created a simulated set of magnitudes, one for each of the stars in the *Almagest*, where each magnitude is that predicted by the model (Equation A4) plus a Gaussian random offset. This data was created for an assumed latitude of 31.2° and extinction coefficient of 0.25 mag/airmass. This is a simulation of what Ptolemy would have observed. Plot vertical axis of the plot (m-V) is the perceived dimming of the starlight when viewed at culmination when compared to stars around zenith. Importantly, we see a distinct upturn to the left, where the most southerly stars are always seen dimmed compared to high-up stars. This upturn to the left is the hallmark of atmospheric extinction, and this is the key property for this paper. With such curves, the latitude of the observer can be determined to better than a degree or so.

# APPENDIX 3: ARE THE MAGNITUDES IN THE THREE CATALOGUES INDEPENDENT?

A key issue for understanding the catalogues of Al Sufi and Tycho is whether their reported magnitudes are independent from earlier catalogues. That is, if Tycho or Al Sufi largely copied the magnitudes from an earlier catalogue, then we might learn nothing about Tycho or Al Sufi's knowledge of extinction. After all, Ulugh Begh copied the magnitudes from the earlier star catalogue of Al Sufi, while Al Sufi copied the positions from the Almagest. Neither Tycho nor Al Sufi would have access to any earlier information on magnitudes beyond that which is contained in the surviving *Almagest*, so any copying or reprocessing would have to derive from the same information that we currently have.

At first look, all of the catalogues are independent, with the evidence being simply that they all quoted magnitudes with different scales and different depths. Tycho's 1004-star catalogue reported magnitudes quantized to the nearest integer, while the other catalogues report magnitudes with three divisions per unit magnitude. Al Sufi reported stars to halfway between 6 and 7, while the other catalogues reported nothing fainter than 6 mag.

At second look, all of the catalogues are independent, with the evidence being that large numbers of stars are reported with different magnitudes. So the magnitudes were not simply copied from one catalogue to the next.

Nevertheless, copying might have been done on a smaller scale, with only some fraction copied. Copying might have been done as straight duplication from the *Almagest* without ever having looked at the sky. Alternatively, copying might have come from some bias caused by the observer knowing the *Almagest* magnitude and this influenced his real observation. This influence is easy to imagine, for example in the cases where the observer's judgment of magnitude places it near the border between two magnitude bins and so the observer (consciously or unconsciously) chooses the bin as given by Ptolemy. In either case, the result would be the same, where some fraction of the reported magnitudes are identical to those in the *Almagest*. The trouble is that there is another expected mechanism that will also produce a large fraction of identical magnitudes, and that is that the two observers both made correct estimates of the same star of the same brightness. So the real question for this appendix is to distinguish the identical reported magnitudes as caused by copying (or any type of influence) from independent measures.

A closer look is needed, and this can only be a detailed comparison of the magnitudes reported by each catalogue. For example, the star Sirius is reported to be first magnitude in all catalogues. But this does not imply copying simply because the star is so bright that anyone would place it in the brightest star category whether or not they copied the magnitude. For another example, Kochab (β UMi) is listed as magnitude 2 in all three catalogues. But this does not imply copying because the star really is second magnitude (V=2.08) so independent observers would likely all report the same magnitude. For a third example, the star 69 Her is listed as magnitude 4, 3, and 5, by Ptolemy, Al Sufi, and Tycho respectively. But, by itself, this does not prove the lack of copying, because we can reasonably invoke some small number of copying errors or variable stars. From these examples and considerations, we see that we have to take a more systematic approach than to simply compare magnitudes reported for single stars.

A bulk comparison of magnitudes can be made from a tabulation of all magnitude combinations for individual stars between a pair of catalogues. Table A5 is such a tabulation for comparing the magnitudes reported by Ptolemy and Al Sufi. Listed are the number of stars that are reported in each magnitude bin for Ptolemy and Al Sufi. To provide a specific reading from the table, in the first line, we see that 12 stars are reported as first magnitude by both Ptolemy and Al Sufi, while one star that Ptolemy labeled as "<1" is simply called first magnitude by Al Sufi, while there are no other stars labeled as first magnitude by Al Sufi. To help guide the eye in the table, the numbers are typed in bold face for the bins were the Ptolemy and Al Sufi magnitudes nominally agree. We see that most of the stars fall along the diagonal in the table, which is to say that both Ptolemy and Al Sufi report identical or similar magnitudes for most stars. This table has 1010 stars. Of these, 559 are reported with identical magnitudes, for 56% of the total. Similarly, 46% of Tycho's stars have the same reported magnitude as in the *Almagest*.

The nearly half with identical magnitudes could be caused by some stars having their magnitudes being copied from the *Almagest*. Alternatively, the reported magnitudes could be identical simply because both Ptolemy and Al Sufi are correctly measuring the same brightness for the individual stars. If all (or most) of the stars had identical magnitudes, then this would make for a good case of copying. But if only roughly half of the stars have identical magnitudes, then we have already disproven the idea that Al Sufi and Tycho simply copied Ptolemy's magnitudes, or did any sort of simple transformation.

Perhaps the best case for independent magnitudes comes from the stars that have greatly discrepant reported magnitudes. In the table for Al Sufi, we see 162 stars (16%) are more than one category apart (so the magnitude difference is roughly >0.6 mag). This fraction is too large to attribute to stellar variability or scribal errors.

We could model the situation by saying that Al Sufi (or Tycho) directly copied some fraction of the magnitudes and made independent magnitude measures for the rest. The copy fraction ($f_{copy}$) must be substantially less than 56%, as many of the star with identical reported magnitudes must have been caused simply by Al Sufi making an independent measure that happened to agree with Ptolemy. If we assume that Al Sufi made the same magnitude estimate as Ptolemy one-third of the time, then $f_{copy}$~1/3 so as to produce the 56% fraction with identical reported magnitudes. If we have the situation where Al Sufi independently reports the same magnitude as Ptolemy 56% of the time, then $f_{copy}$=0. So to know whether the copy fraction is large or small will take a detailed analysis.

We can quantitatively measure $f_{copy}$ by looking at stars whose modern magnitude is within a narrow range. In particular, we will look at four sets of stars; those with 1.9<V<2.1, 2.9<V<3.1, 3.9<V<4.1, and 4.9<V<5.1. Within such sets, we can be sure that the stars were presented to both observers essentially identically, and that there can be no effects caused by one observer happening to have more or less stars near the top or bottom of the range. We have constructed matrices (tables) with the stars of each set, where each element (entry) is the number of stars reported in each magnitude range for two observers. An example matrix is presented in Table A8 for stars with 3.9<V<4.1 as reported by Tycho and Ptolemy. The total number of stars in the matrix is N=58. Out of all these stars, $f_{T3}$, $f_{T4}$, and $f_{T5}$ are the fractions that were reported by Tycho to be with m=3, m=4, and m=5 respectively. The fractions of these stars reported by Ptolemy to be m~3, m~4, and m~5 are denoted by $f_{P3}$, $f_{P4}$, and $f_{P5}$ respectively. For this example, I have combined stars reported as being just brighter and fainter along with the stars of the central magnitude.

The distribution of numbers will depend on the copying fraction ($f_{copy}$). If Tycho simply copied all of his reported magnitudes ($f_{copy}$=1), then all of the off-diagonal elements must be zero, while the diagonal elements will simply equal to the fractions for Ptolemy. Alternatively, if Tycho made his magnitude estimates completely independent from the *Almagest* ($f_{copy}$=0), then the numbers in each element should be the products of N and the two fractions from Tycho and Ptolemy. If Tycho copied some fraction of the *Almagest* magnitudes, then the model predicted numbers for each element are a linear combination of the two extremes. Table A9 gives the model predictions for each element. The uncertainty in each element is the usual binomial uncertainty of $[NP(1-P)]^{0.5}$, where P is the probability for calculating each element. This model has only one free parameter, $f_{copy}$, which determines the relative size of the off-diagonal elements.

We now have an observed matrix and a model predicted matrix (along with the uncertainty). These can be compared in the usual chi-square calculation. That is, each matrix element has a ratio formed that is the difference between the observed and model numbers divided by the uncertainty. The chi-square equals the sum (over all matrix elements) of the squares of these ratios. This chi-square will be smallest for the best fit value of $f_{copy}$. That is, the optimal value of $f_{copy}$ can be found by varying its value until the chi-square is minimized. The number of degrees of freedom are the number of non-zero model elements minus unity. The reduced chi-square equals the minimized chi-square divided by the number of degrees of freedom. The model produces a reasonable fit to the data when the reduced chi-square is roughly equal to unity. The one-sigma error bars are for the range of $f_{copy}$ where the chi-square is within 1.0 of the minimum, the two-sigma error bars are for when the chi-square remains within 4.0 of the minimum, while the three-sigma error bars are for when the chi-square is within 9.0 of minimum. With this, we have a standard way of calculating the copy fraction (and its uncertainty) in a quantitative and authoritative way as derived from simple counts of reported magnitudes.

Table A10 summarizes the chi-square fits and the resultant $f_{copy}$ values. For the first line (for testing the N=15 stars with 1.9<V<2.1 common to the catalogues of Tycho and the *Almagest*), we see that the best fit $f_{copy}$ is 0.08±0.15, which is to say that the copying fraction is low and consistent with zero. The $f_{copy}$=0 hypothesis has a chi-square that is only 0.6 above the minimum, with this indicating that the one-sigma confidence region includes zero copying. For Tycho's stars, we see a low copying fraction, typically around 10%, with most being consistent with zero. For the 3.9<V<4.1 stars, the $f_{copy}$=0 hypothesis is rejected at the three-sigma confidence level, because the chi-square is more than 9.0 above the minimum.

### A3.1. *Al Sufi's Catalogue*

The first test for copying is simply to look at the fraction of stars with identical reported magnitudes. For one comparison, we can look at stars that have identical reported values (i.e., the bold face diagonal entries in Table A5). Here, I am equating "1-2" for Al Sufi with "<1" for Ptolemy, "2-1" for Al Sufi with ">2" for Ptolemy, "2-3" for Al Sufi with "<2" for Ptolemy, and so on. For these comparisons with 1/3-magnitude bins, 55% of Al Sufi's reported magnitudes are identical to those reported by Ptolemy. For another comparison, we can look at stars reported to have the same magnitudes to within one magnitude bins, where ">2", "2", and "<2" are lumped together, and so on. With these 1-magnitude bins, 84% of Al Sufi's magnitudes are the same as those of Ptolemy. These ratios (55% and 84%) present a strict upper limit on the copying fraction. Indeed, the real $f_{copy}$ must be substantially lower, because the non-copied fraction must have many stars reported to have the same magnitudes simply from independent measures both reporting the correct value.

The second test for copying is to look at the four sets of nearly-identical brightness stars (e.g., 1.9<V<2.1) as described earlier. We can proceed with two types of binning. The first is to lump all the magnitudes (for both Al Sufi and Ptolemy) into bins nominally stretching 1 mag in breadth. The second is to keep the bins as reported (with roughly 1/3 magnitude bin sizes) and to equate bin names as above. With four sets of stars and two bin widths each, we have eight chi-square fits, with the results presented in Table A10. For the one-mag bin sizes, the average $f_{copy}$ is close to 62% while the possibility of $f_{copy}$=0 is strongly rejected. For the 1/3 mag bin sizes, $f_{copy}$ varies from 6% up to 63%, with the possibility of $f_{copy}$=0 strongly rejected for three of the four magnitude ranges. The copy fraction appears to be the highest for the bright stars and to have dropped substantially for fainter stars. The copy fraction is systematically lower for the 1/3 mag bin sizes as compared to the 1 mag bin sizes. This means that Al Sufi copied something like a third of the magnitudes directly, while another third were influenced to have the same numerical digit. For example, if Ptolemy quoted a magnitude of "<3", then Al Sufi would repeat "3-4" a third of the time, would give "3" or "3-2" another third of the time, and would pick any other magnitude the last third of the time. For the times when Al Sufi made changes from Ptolemy, he was influenced such that he had a strong tendency to just report a small shuffle within the same 1-mag bin. So we see that Al Sufi made a combination of something like straight copying plus significant influence of the prior report.

   A third test is to look to see if Al Sufi has repeated any large discrepancies that Ptolemy reported. If both catalogues report the same roughly-correct magnitude then this has easy interpretations under both the hypotheses of copying and independent measures. But if Ptolemy reports a magnitude that is widely different than the real brightness and if Al Sufi reports the identical error, then this would be good evidence for copying. For this, we must compare the reported magnitudes on the modern system (m) to the modern magnitudes (V) so as to spot a discrepancy. The discrepancy is notated as $\Delta m = m - V$. A star with $\Delta m < -1$ is one for which the old report places the star much fainter than it really is, while $\Delta m > 1$ is for a star whose old magnitude is much brighter than the real brightness of the star. If discrepant Almagest magnitudes are simply copied by Al Sufi then we would see a similar discrepancy, whereas if Al Sufi made independent observations then we would see the normal scatter of discrepancies for these same stars. We need some criterion to know if the Al Sufi magnitude is to be considered to have a similar discrepancy, and I will arbitrarily and reasonably take a similar discrepancy to be if the Al Sufi $\Delta m$ is more than half as large in magnitude as the discrepancy for Ptolemy. Thus, Sirius (V=−1.40) is reported by Ptolemy to be of first magnitude (m=0.37) for a discrepancy of $\Delta m = 1.77$, while Sirius is also reported by Al Sufi to be of first magnitude (m=0.74) for a discrepancy of $\Delta m = 2.14$, so both catalogues have similar large discrepancies. In this case, the mutual large discrepancy is easily explained by Sirius being so much brighter than the other first magnitude stars. By this criterion, stars with $0.00 < m < 2.00$ have a 60% copying rate, stars with $2.00 < m < 3.00$ have a 64% copying rate, stars with $3.00 < m < 4.00$ have a 42% copying rate, stars with $4.00 < m < 5.00$ have a 37% copying rate, and stars with $5.00 < m$ have a 65% copying rate. Taking all the Almagest high-discrepancy stars together, the copying rate is 49%.

   In all, we know that Al Sufi did not simply copy all of the magnitudes in the Almagest, but some substantial amount of copying (or some equivalent influence) did occur. We can say that Al Sufi copied roughly a third of his magnitudes and was greatly influenced for another third of his magnitude.

   In the above, I have modeled the situation as if Al Sufi had simply blindly copied some fraction of the Almagest magnitudes, but the real situation is likely more complex. It could be that Al Sufi observed all the stars but let Ptolemy's magnitudes (consciously or unconsciously) influence their observations. Or maybe for magnitudes observed to be ambiguous between two bins, Al Sufi let the Almagest decide. Or maybe he simply averaged his own magnitudes with those in the Almagest. At this late time, I see no realistic way to decide the exact nature of the 'copying'.

**A3.2.** *Tycho's Catalogue*

For the first test of the copying fraction, we can tabulate the number of identical magnitudes reported by both Ptolemy and Tycho. For the 1004-star catalogue, the stars that have magnitudes reported by Ptolemy and Tycho, 38% of them are identical. If we allow stars with "<" and ">" to have been combined into one simplified reported bin (e.g., if Tycho had copied all the stars labeled by the Almagest as '>3', '3', and '<3' into one bin labeled '3'), then we have the bold faced entries in Table A6. For this, we have 46% of Tycho's magnitudes being the same as in the *Almagest*. This provides a strict upper limit on the copying fraction. The real value of $f_{copy}$ is inevitably substantially smaller due to the fact that the un-copied fraction will have most of the stars being reported in the same 1-magnitude wide bins. Indeed, the accounting for this uncopied-fraction is at the heart of the second method.

      The second measure of the copying fraction comes from the analysis of the distribution of reported magnitudes for sets of stars within a narrow range of modern magnitudes. An example for Tycho's 58 stars (in his 1004-star catalogue) between V=3.9 and V=4.1 is presented in Table A8, while the model is presented in Table A9. The results of these fits are presented in Table A10. We see that $f_{copy}$ is always low, and consistent with zero for three-out-of-four sets. From these results, I would characterize the copying fraction as being quite low, perhaps with 10% as the typical value.

      For the third measure of $f_{copy}$ for Tycho, we can see whether the discrepant magnitudes reported in the Almagest are reproduced in Tycho's 1004-star catalogue. As with the same test for Al Sufi's catalogue, I will take the discrepant magnitudes in the Almagest to be those for which m-V differs by more than one magnitude from zero, and I will take the Tycho magnitude to be copied if the m-V value for the Tycho magnitude is greater than half that of the Almagest. I find 43 stars with discrepant magnitudes in the *Almagest* which also have magnitudes reported by Tycho. By this criterion, 18 of the stars also have a discrepancy, and all of these have the same sign (positive or negative) as for the *Almagest*. Thus, the copying rate for stars with high discrepancies is roughly 42%. About half of the doubly discrepant stars are very close to Tycho's southern horizon (e.g., ε CMa and σ Sgr), likely variable so the modern magnitude is suspect (i.e., Betelgeuse), or far brighter than its magnitude bin average (i.e., Sirius). The doubly discrepant stars appears to be dominated by a special case of stars close to Tycho's southern horizon. Indeed, these stars will only be occasionally visible to Tycho on good nights, so it is easy to take them as a special case, where he took the expedient of adopting the *Almagest* magnitude because he could not make a good measure of his

own.  The double degeneracy copying rate for stars north of the equator (plus ignoring Betelgeuse) turns out to be 24%.

The three methods point to copying rates between 8% to 24%.  Three of the measures are consistent with zero copying, but two are inconsistent at roughly the three-sigma level.  The best measure of $f_{copy}$ is from the second method, with the typical rate being perhaps 10%.  In all, it looks like Tycho copied few of the magnitudes from the *Almagest*.

The results for Tycho's 777-star catalogue are similar and the conclusions identical as for the 1004-star catalogue.

**TABLE A5. Ptolemy and Al Sufi magnitude distribution.**

|      | 1  | <1 | >2 | 2  | <2 | >3 | 3  | <3 | >4 | 4   | <4 | >5 | 5   | 6  | faint |
|------|----|----|----|----|----|----|----|----|----|-----|----|----|-----|----|-------|
| 1    | **12** | 1  | …  | …  | …  | …  | …  | …  | …  | …   | …  | …  | …   | …  | …     |
| 1-2  | 1  | **1**  | …  | …  | …  | …  | …  | …  | …  | …   | …  | …  | …   | …  | …     |
| 2-1  | …  | …  | **1**  | 1  | …  | …  | …  | …  | …  | …   | …  | …  | …   | …  | …     |
| 2    | …  | …  | 2  | **22** | 1  | 1  | …  | …  | …  | …   | …  | …  | …   | …  | …     |
| 2-3  | …  | …  | …  | 1  | **4**  | …  | 2  | …  | …  | …   | …  | …  | …   | …  | …     |
| 3-2  | …  | …  | …  | 4  | …  | **4**  | 3  | …  | …  | …   | …  | …  | …   | …  | …     |
| 3    | …  | …  | …  | 5  | 1  | 4  | **82** | 2  | 1  | …   | …  | …  | …   | …  | …     |
| 3-4  | …  | …  | …  | …  | …  | 3  | 66 | **13** | 7  | 6   | …  | …  | …   | …  | …     |
| 4-3  | …  | …  | …  | …  | …  | …  | 9  | 1  | **30** | 34  | 1  | …  | …   | …  | …     |
| 4    | …  | …  | …  | 1  | …  | …  | 7  | 1  | 33 | **212** | 2  | …  | 4   | …  | 2     |
| 4-5  | …  | …  | …  | 1  | 1  | …  | 3  | …  | 4  | 62  | **12** | …  | 2   | …  | …     |
| 5-4  | …  | …  | …  | …  | …  | …  | …  | …  | 1  | 8   | …  | …  | 11  | …  | 1     |
| 5    | …  | …  | …  | …  | …  | …  | …  | 1  | 4  | 31  | …  | 6  | **119** | 1  | 4     |
| 5-6  | …  | …  | …  | …  | …  | …  | …  | …  | 3  | 12  | 2  | 2  | 45  | 1  | 2     |
| 6-5  | …  | …  | …  | …  | …  | …  | …  | …  | …  | …   | …  | …  | 1   | …  | …     |
| 6    | …  | …  | …  | …  | …  | …  | 1  | …  | 1  | 3   | 1  | 1  | 20  | **39** | 4     |
| 6-7  | …  | …  | …  | …  | …  | …  | …  | …  | …  | …   | …  | …  | 2   | …  | **5**     |

**TABLE A6. Ptolemy and Tycho (1004-star catalogue) magnitude distribution.**

|   | 1 | <1 | >2 | 2  | <2 | >3 | 3  | <3 | >4 | 4   | <4 | >5 | 5  | 6  | faint |
|---|---|----|----|----|----|----|----|----|----|-----|----|----|----|----|-------|
| 1 | **9** | 1  | …  | 2  | …  | …  | …  | …  | …  | …   | …  | …  | …  | …  | …     |
| 2 | 1 | 1  | **3**  | **18** | 5  | 1  | 12 | …  | …  | …   | …  | …  | …  | …  | …     |
| 3 | … | …  | …  | 2  | …  | **8**  | **95** | 13 | 13 | 23  | 1  | …  | …  | …  | 1     |
| 4 | … | …  | …  | …  | …  | …  | 33 | 4  | **40** | **146** | 8  | …  | 27 | 2  | 9     |
| 5 | … | …  | …  | …  | …  | …  | 2  | …  | 4  | 85  | 5  | **2**  | **64** | 3  | …     |
| 6 | … | …  | …  | …  | …  | …  | …  | 1  | …  | 16  | 3  | …  | 40 | **32** | …     |

**TABLE A7. Ptolemy and Tycho (777-star catalogue) magnitude distribution.**

|    | 1 | <1 | >2 | 2 | <2 | >3 | 3 | <3 | >4 | 4 | <4 | >5 | 5 | 6 | faint |
|----|---|----|----|---|----|----|---|----|----|---|----|----|---|---|-------|
| 1  | **7** | **...** | … | 1 | … | … | … | … | … | … | … | … | … | … | … |
| 1. | **2** | **1** | … | 1 | … | … | … | … | … | … | … | … | … | … | … |
| 2: | 1 | 1 | **2** | 1 | **...** | … | … | … | … | … | … | … | … | … | … |
| 2  | … | … | **...** | **14** | **5** | 1 | 9 | … | … | … | … | … | … | … | … |
| 2. | … | … | **1** | **3** | **...** | … | 3 | … | … | … | … | … | … | … | … |
| 3: | … | … | … | … | … | **1** | **11** | **...** | … | 1 | … | … | … | … | … |
| 3  | … | … | … | 2 | … | **6** | **63** | **12** | … | 9 | … | … | 1 | … | … |
| 3. | … | … | … | … | … | **...** | **14** | **1** | 6 | 9 | … | … | … | … | 1 |
| 4: | … | … | … | … | … | … | 9 | 1 | **5** | **4** | **1** | … | 1 | … | … |
| 4  | … | … | … | … | … | … | 21 | 3 | **23** | **98** | **2** | … | 13 | 1 | 5 |
| 4. | … | … | … | … | … | … | 3 | … | **9** | **29** | **...** | … | 7 | … | 2 |
| 5: | … | … | … | … | … | … | 1 | … | 2 | 16 | 1 | **1** | **3** | 6 | … |
| 5  | … | … | … | … | … | … | 1 | … | 1 | 56 | 4 | **...** | **34** | 1 | … |
| 5. | … | … | … | … | … | … | … | … | … | 11 | … | **...** | **12** | … | … |
| 6: | … | … | … | … | … | … | … | … | … | 2 | … | … | 8 | **1** | … |
| 6  | … | … | … | … | … | … | … | 1 | … | 9 | 4 | … | 22 | **23** | … |
| 6. | … | … | … | … | … | … | … | … | … | 4 | … | … | 5 | **8** | … |

## TABLE A8. 3.9<V<4.1 stars as reported by Tycho and Ptolemy

|  | Ptolemy: m~3 | Ptolemy: m~4 | Ptolemy: m~5 | |
|---|---|---|---|---|
| Tycho: m=3 | 7 stars | 4 stars | 0 stars | $f_{T3}=0.19$ |
| Tycho: m=4 | 5 stars | 26 stars | 2 stars | $f_{T4}=0.57$ |
| Tycho: m=5 | 0 stars | 12 stars | 2 stars | $f_{T5}=0.24$ |
|  | $f_{P3}=0.21$ | $f_{P4}=0.72$ | $f_{P5}=0.07$ | N=58 stars |

## TABLE A9. Model predicted numbers

|  | Ptolemy: m~3 | Ptolemy: m~4 | Ptolemy: m~5 |
|---|---|---|---|
| Tycho: m=3 | $Nf_{P3}f_{T3}(1-f_{copy})$ $+Nf_{P3}f_{copy}$ | $Nf_{P4}f_{T3}(1-f_{copy})$ | $Nf_{P5}f_{T3}(1-f_{copy})$ |
| Tycho: m=4 | $Nf_{P3}f_{T4}(1-f_{copy})$ | $Nf_{P4}f_{T4}(1-f_{copy})$ $+Nf_{P4}f_{copy}$ | $Nf_{P5}f_{T4}(1-f_{copy})$ |
| Tycho: m=5 | $Nf_{P3}f_{T5}(1-f_{copy})$ | $Nf_{P4}f_{T5}(1-f_{copy})$ | $Nf_{P5}f_{T5}(1-f_{copy})$ $+Nf_{P5}f_{copy}$ |

**TABLE A10. Copying fractions for Al Sufi and Tycho**

| Catalogue | Source? | Stars | Bin size | N | $f_{copy}$ | $\chi^2$(best) | $\chi^2(f_{copy}=0)$ |
|---|---|---|---|---|---|---|---|
| Al Sufi | *Almagest* | 1.9<V<2.1 | 1 mag | 18 | 0.84 ± 0.15 | 0.8 | 18.2 |
| Al Sufi | *Almagest* | 2.9<V<3.1 | 1 mag | 36 | 0.62 ± 0.14 | 5.3 | 60.8 |
| Al Sufi | *Almagest* | 3.9<V<4.1 | 1 mag | 82 | 0.62 ± 0.08 | 12.0 | 103.8 |
| Al Sufi | *Almagest* | 4.9<V<5.1 | 1 mag | 64 | 0.44 ± 0.14 | 14.2 | 51.6 |
| Al Sufi | *Almagest* | 1.9<V<2.1 | 1/3 mag | 18 | 0.63 ± 0.14 | 5.9 | 34.1 |
| Al Sufi | *Almagest* | 2.9<V<3.1 | 1/3 mag | 36 | 0.33 ± 0.07 | 18.8 | 94.8 |
| Al Sufi | *Almagest* | 3.9<V<4.1 | 1/3 mag | 82 | 0.06 ± 0.03 | 137.4 | 142.8 |
| Al Sufi | *Almagest* | 4.9<V<5.1 | 1/3 mag | 64 | 0.19 ± 0.06 | 67.6 | 85.8 |
| Tycho-1004 | *Almagest* | 1.9<V<2.1 | 1 mag | 15 | 0.08 ± 0.15 | 3.3 | 3.9 |
| Tycho-1004 | *Almagest* | 2.9<V<3.1 | 1 mag | 27 | 0.09 ± 0.15 | 6.5 | 6.9 |
| Tycho-1004 | *Almagest* | 3.9<V<4.1 | 1 mag | 58 | 0.24 ± 0.08 | 8.2 | 18.8 |
| Tycho-1004 | *Almagest* | 4.9<V<5.1 | 1 mag | 53 | 0.11 ± 0.08 | 3.9 | 6.1 |
| Tycho-777 | *Almagest* | 1.9<V<2.1 | 1 mag | 15 | 0.08 ± 0.15 | 3.3 | 3.9 |
| Tycho-777 | *Almagest* | 2.9<V<3.1 | 1 mag | 28 | 0.09 ± 0.15 | 6.8 | 7.2 |
| Tycho-777 | *Almagest* | 3.9<V<4.1 | 1 mag | 49 | 0.21 ± 0.10 | 6.8 | 13.9 |
| Tycho-777 | *Almagest* | 4.9<V<5.1 | 1 mag | 44 | 0.12 ± 0.09 | 7.8 | 10.2 |
| Tycho-777 | *Almagest* | 1.9<V<2.1 | 1/3 mag | 15 | <0.00 | ... | 15.9 |
| Tycho-777 | *Almagest* | 2.9<V<3.1 | 1/3 mag | 28 | <0.00 | ... | 46.5 |
| Tycho-777 | *Almagest* | 3.9<V<4.1 | 1/3 mag | 49 | 0.13 ± 0.06 | 32.1 | 41.3 |
| Tycho-777 | *Almagest* | 4.9<V<5.1 | 1/3 mag | 44 | 0.09 ± 0.07 | 22.7 | 25.0 |

# APPENDIX 4: MODERN EXPERIMENTS

So the pre-telescopic star catalogues were somehow corrected for extinction, but many practical and important questions remain unanswered. Is it possible to correct for extinction without modern equipment? What methods might have been employed? How accurate can those methods be? How far south will stars be recorded? Are there any operating psychological or physiological effects based on proximity to the horizon? Would the observed accuracy of the corrections in the old catalogues require fanatical attention to details or mere casual attention? Plausible answers to these questions can be made from modern experiments, where people step outside, look at the stars on clear nights, and try for themselves to correct for extinction. We will not fool ourselves into thinking that modern experimentation will reproduce the exact or even approximate procedures of the old observers, but modern experiments can well show what is possible and what is reasonable.

      A hallmark of my research in astronomical history has been to create modern 'ground truth' and to experimentally test theoretical models of celestial visibility. These modern experiments have been based on extensive series of observations by myself, the recreation of entire ancient databases, and vast public observing programs[35]. The programs have already answered the question "How far south will stars be recorded?", with the answer being that only first magnitude stars will be recorded amongst those that culminate within ~5° of the southern horizon[36]. The point being that real experience often shows realities and key insights and problems that would be entirely missed by an 'armchair historian'. With this background, it is natural to perform modern experiments to see how well simple observations can correct for extinction.

      On 5 nights from 21 May to 12 June 2011, I made a series of magnitude estimates from very dark sites in Arizona, New Mexico, and Texas (with an average latitude of 32.6°). The observations were designed to measure and correct extinction, all while using zero equipment. I intentionally took a casual attitude towards the measures, while *not* making repeated measures or careful determination of horizons or altitudes. The basic idea was to calibrate extinction by watching the brightness of stars as they rise and set. This fading of brightness was measured as a function of the star's altitude, which was quantified by the apparent height above the horizon as measured in finger-widths (held at arm's length). This calibrated dimming was then applied as a correction for measured magnitudes as a function of the observed altitude of the star.

      All stellar photometry must start with the definition of the zero magnitude and the adoption of standard stars of known magnitude. For pre-telescopic observers measuring magnitudes, they must have adopted some equivalent procedure, even though it would likely be substantially less formal and precise. For my modern experiment, I adopted ~5 stars far from the horizon as standards, with my assumed magnitudes taken from their modern V magnitudes. For example, for June evenings, I assumed m=0.0 for

Vega, m=2.1 for Rasalhague (α Oph), m=2.8 for Vindemiatrix (ε Vir), m=3.8 for ζ Boo, and m=4.5 for ι Ser.  Other star magnitudes were then made by direct interpolation of perceived brightness between two of the standard stars.  Long experience by a very large community of amateur variable star observers and by myself show that such magnitudes can be estimated with an one-sigma accuracy of 0.3 mag or better.  Magnitudes were recorded to a precision of 0.1 mag.

      Stars near the horizon had their magnitudes directly compared to stars far from the horizon.  Sometimes the comparison stars were standard stars near the zenith, while sometimes the comparison stars were secondary standards (themselves calibrated that night from the primary standards near zenith) with a similar azimuth and at an altitude of perhaps 45° (so that the stars compared are easily in the same view).  Importantly, the dimming from extinction is identical for whether the low star is directly compared to a standard at zenith or compared to a chain of secondary stars stretching from low altitude to the zenith.  The natural procedure is to keep glancing back and forth between the stars as well as to stare at their midpoint.

      For each magnitude estimate, I also estimated the altitude of the star above the horizon.  This estimate was made with the fast and easy method of holding up my extended arm, placing the bottom my hand on a level with my eye, and counting the number of fingers that separate the star from the horizon.  This method likely has an accuracy of only 30%, because the horizon was usually not seen (so we are relying on my evaluation of the horizontal direction), because the width of the fingers vary (both with the number of fingers used and how close to the fingertips is used), and because for more than four fingers I simply remembered the top of the fourth finger for when I moved my hand.  All of this was intentionally casual, and we can think of many ways to improve the accuracy.

      I watched many stars rise and set, measuring the magnitude (with respect to standard stars far from the horizon) and altitude (in finger widths).  For each of these 93 observations, I also measured the star's magnitude when it was far from the horizon.  With this, I can determine the dimming (in magnitudes) for each observation.  A plot of the dimming (in magnitudes) versus altitude (in finger widths) then shows the systematic effect of extinction, with a characteristic scatter of ~0.3 mag.  Intentionally, I lumped all of the data into a single plot, even though the modern photometrist knows that the extinction coefficient changes substantially from night-to-night and site-to-site.  I then drew a crude curve down the middle of the observations, and this became my extinction correction.

      I also made a separate series of observations of stars around their culmination for the purpose of making a small catalogue of magnitudes.  These stars ranged widely in declination, with many near the southern horizon.  For each observation, I recorded the star's altitude (in finger widths) and then applied my crude extinction correction.  To simulate the quantization of magnitudes due to the binning, I then took the extinction

corrected magnitudes and rounded them to the nearest integer. The result is a modern magnitude catalogue of 176 stars with extinction corrected magnitudes quoted only to the nearest integer. I also created a catalogue of quantized magnitudes that did not have the extinction correction. These magnitudes are plotted versus the star declination in Figures A3 and A4.

The plot in Figure A3 is for the no-correction case, and these data can be used in a chi-square fit. I find the best fit latitude to be $32.1° ± 0.6°$ and the best fit extinction coefficient to be $0.28 ± 0.02$ mag/airmass. This extinction is typical of good nights at low altitudes in the legendary clear skies of the American southwest. With a small catalogue of casual magnitudes, I have the latitude correct with an error bar of 40 miles in the north/south direction. This experience provides proof that the basic method can indeed determine the observer's latitude -- provided that no extinction correction was applied.

For the plot of the extinction corrected magnitudes (see Figure A4), we see a nearly flat m-V as a function of declination, although the most southerly are mostly positive. A formal chi-square fit produces an extinction coefficient of 0.07 mag/airmass with a one-sigma uncertainty of ±0.02. With the real extinction being 0.28 mag/airmass, this shows that my extinction corrections only got ~3/4 of the correction made, or that the accuracy of my correction was only 25%. The imperfection of my empirical equipment-less correction is easily attributed to the purposefully casual nature of the observations. The conclusion is that it is easy to correct for extinction to ~25% accuracy, even with no equipment, no modern knowledge, and no particular care taken.

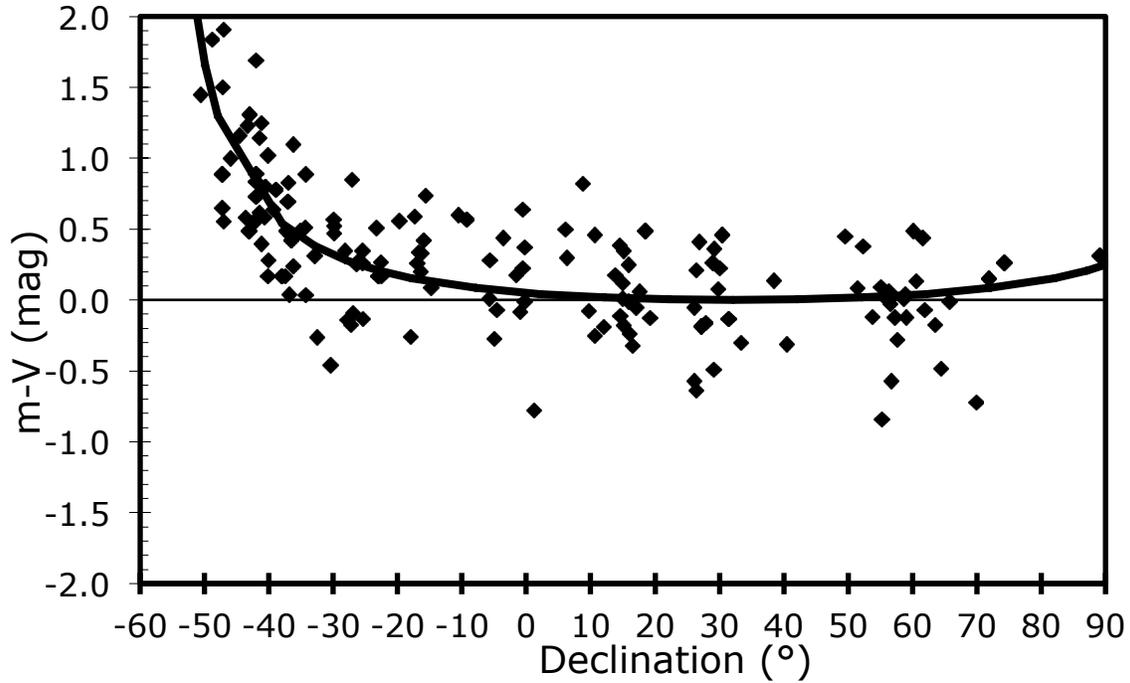

**Fig. A3. Modern magnitudes without extinction correction.**
I have observed magnitudes for a small catalogue of 176 stars as viewed on 5 nights in the summer of 2011 from the American southwest. The observed magnitudes were taken from interpolation of the brightness between standard stars far from the horizon, and then the values were rounded to the nearest integer (to simulate the binning of the catalogued magnitudes). These magnitudes show the expected rise in m-V towards the left, which is to say that the most southerly stars at culmination are substantially dimmed. A formal chi-square fit (with the best fit model displayed as the thick curved line) gives a latitude of $\lambda=32.1° \pm 0.6°$ and an extinction coefficient of k=0.28 ± 0.02 mag/airmass. This demonstrates that the method can provide correct and accurate measures of the latitude.

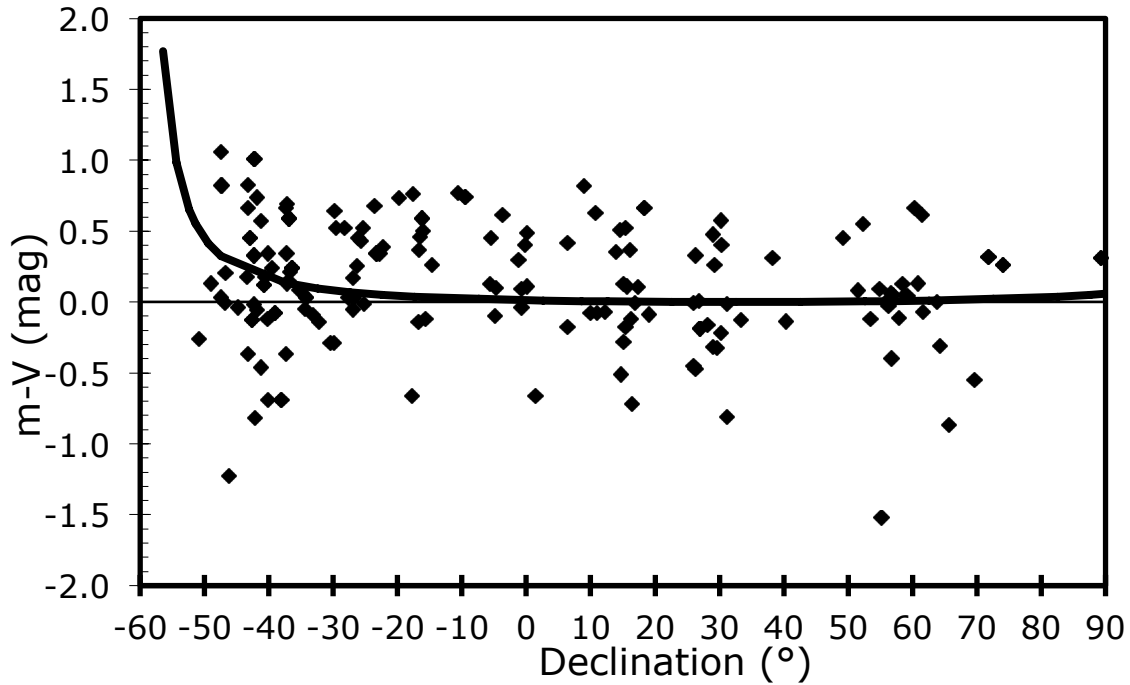

**Fig. A4. Modern magnitudes with extinction correction.**
The magnitudes in Figure A3 were also corrected for extinction. Indeed, this figure shows no obvious rise towards the left, although the southern stars do have a majority with positive m-V. A formal chi-square analysis (with the best fit model given by the thick curve) gives a derived value of k=0.07±0.02 mag/airmass. With this imperfect extinction correction, roughly 3/4 of the atmospheric dimming has been compensated. This modern experiment demonstrates that it is easy to correct for the extinction to ~25% accuracy, despite using no equipment, no modern knowledge, and an intentionally casual manner.